\documentclass[twocolumn]{aastex62}

\usepackage{enumitem}

\newcommand{\mfive}{M$_{500}$}
\newcommand{\chandra}{{\it Chandra}}
\newcommand{\xmm}{{\it XMM-Newton}}
\newcommand{\planck}{{\it Planck}}
\newcommand{\spire}{{\it Herschel}--SPIRE}
\newcommand{\hst}{{\it HST}}
\newcommand{\hitomi}{{\it Hitomi}}

\newcommand{\mjysr}{MJy~sr$^{-1}$}
\newcommand{\eg}{{\it e.g.}}
\newcommand{\ie}{{\it i.e.}}
\newcommand{\taue}{$\tau_{\textrm{\footnotesize e}}$}
\newcommand{\vlos}{$v_{z}$}
\newcommand{\radius}{$R_{2500}$}
\newcommand{\kms}{km~s$^{-1}$}
\newcommand{\ymap}{$y$--map}
\newcommand{\te}{$T_{\textrm{\footnotesize e}}$}
\newcommand{\asix}{Abell 0697}
\newcommand{\aeight}{Abell 1835}
\newcommand{\roone}{RX J0152.7$-$1357}
\newcommand{\rtwo}{RX J1226.9$+$3332}
\newcommand{\moone}{MACS J0018.5$+$1626}
\newcommand{\motwo}{MACS J0025.4$-$1222}
\newcommand{\mfour}{MACS J0454.1$-$0300}
\newcommand{\mseven}{MACS J0717.5$+$3745}
\newcommand{\rthree}{RX J1347.5$-$1145}
\newcommand{\mtwo}{MACS J2129.4$-$0741}

\submitjournal{ApJ}
\shorttitle{Internal and Bulk Cluster Velocities}
\shortauthors{Sayers et al.}

\begin{document}

\title{Imaging the Thermal and Kinematic Sunyaev-Zel'dovich Effect Signals in a Sample of
Ten Massive Galaxy Clusters: Constraints on Internal Velocity Structures and Bulk Velocities}

\correspondingauthor{Jack Sayers}
\email{jack@caltech.edu}

\author[0000-0002-8213-3784]{Jack Sayers}
\affil{California Institute of Technology, 1200 E. California Blvd., MC 367-17, 
Pasadena, CA 91125, USA}

\author{Alfredo Monta{\~n}a}
\affil{Consejo Nacional de Ciencia y Tecnolog{\'i}a-Instituto Nacional
de Astrof{\'i}sica, {\'O}ptica, y Electr{\'o}nica (CONACyT-INAOE), 
Luis Enrique Erro 1, 72840 Puebla, Mexico}

\author{Tony Mroczkowski}
\affil{European Southern Observatory (ESO), 
Karl-Schwarzschild-Str. 2, D-85748 Garching b. M{\"u}nchen, Germany}

\author{Grant W. Wilson}
\affil{University of Massachusetts, Amherst, MA 01003, USA}

\author{Michael Zemcov}
\affil{Rochester Institute of Technology, Rochester, NY 14623, USA}
\affil{Jet Propulsion Laboratory, 4800 Oak Grove Dr., Pasadena, CA 91109, USA}

\author{Adi Zitrin}
\affil{Ben-Gurion University of the Negev, P.O. Box 653, Be'er-Sheva 8410501, Israel}

\author{Nath{\'a}lia Cibirka}
\affil{Ben-Gurion University of the Negev, P.O. Box 653, Be'er-Sheva 8410501, Israel}

\author{Sunil Golwala}
\affil{California Institute of Technology, 1200 E. California Blvd., MC 367-17, 
Pasadena, CA 91125, USA}

\author{David Hughes}
\affil{Instituto Nacional
de Astrof{\'i}sica, {\'O}ptica, y Electr{\'o}nica (INAOE), 
Aptdo. Postal 51 y 216, 7200, Puebla, Mexico}

\author{Daisuke Nagai}
\affil{Department of Physics, Yale University, New Haven, CT 06520, USA}
\affil{Yale Center for Astronomy and Astrophysics, Yale University, New Haven, CT 06520, USA}

\author{Erik D. Reese}
\affil{Moorpark College, 7075 Campus Rd., Moorpark, CA 93021, USA}

\author{David S{\'a}nchez}
\affil{Consejo Nacional de Ciencia y Tecnolog{\'i}a-Instituto Nacional
de Astrof{\'i}sica, {\'O}ptica, y Electr{\'o}nica (CONACyT-INAOE), 
Luis Enrique Erro 1, 72840 Puebla, Mexico}

\author{John Zuhone}
\affil{Harvard-Smithsonian Center for Astrophysics, 60 Garden St., Cambridge, MA 02138, USA}

\begin{abstract}

We have imaged the Sunyaev-Zel'dovich (SZ) effect signals at 140 and 270~GHz
towards ten galaxy clusters with Bolocam and AzTEC/ASTE.
We also used \planck\ data to constrain the signal at large angular scales,
\spire\ images to subtract the brightest galaxies
that comprise the cosmic infrared background (CIB), \chandra\ imaging
to map the electron temperature \te\ of the intra-cluster medium (ICM), and \hst\
imaging to derive models of each galaxy cluster's mass density.
The galaxy clusters gravitationally lens the background CIB, which produced an on-average 
reduction in brightness towards the galaxy clusters' centers after the brightest galaxies
were subtracted. We corrected for this deficit, which was between
5--25\% of the 270~GHz SZ effect signal within \radius.
Using the SZ effect measurements, along with the X-ray constraint on \te,
we measured each galaxy cluster's average line of sight (LOS) velocity \vlos\ within 
\radius, with a median per-cluster uncertainty of $\pm$700~\kms.
We found an ensemble-mean $\langle$\vlos$\rangle$ of $430 \pm 210$~\kms,
and an intrinsic cluster-to-cluster scatter $\sigma_{\textrm{\footnotesize int}}$
of $470 \pm 340$~\kms. 
We also obtained maps of \vlos\ over each galaxy cluster's face with an angular
resolution of 70\arcsec.
All four galaxy clusters previously identified as having a merger oriented along the
LOS showed
an excess variance in these maps at a significance of $\simeq 2$--$4 \sigma$,
indicating an internal \vlos\ rms of $\gtrsim 1000$~\kms.
None of the six galaxy clusters previously identified as relaxed or plane of sky mergers
showed any such excess variance.

\end{abstract}


\section{Introduction} \label{sec:intro}

Velocity measurements have long been used to probe the detailed properties
of large-scale structure, for example the velocity dispersion of 
cluster galaxies \citep{Zwicky1937} and the galaxy rotation curves that
provided evidence of dark matter \citep{Rubin1980}.
As another more recent example, the \hitomi\ X-ray satellite
provided the first direct measurement of the velocity structure
of the intra-cluster medium (ICM) in the core of the 
Perseus cluster \citep{Hitomi2016,Hitomi2018}, providing new
insights on the interaction between the ICM and the central
active galactic nucleus as well as a large-scale velocity
shear due to cosmic accretion and mergers
\citep{Lau2017,Zuhone2018}.
In addition, the statistical properties of the cosmological velocity
field can be used to constrain a range of parameters, particularly those
related to dark energy and possible modifications of general
relativity (\eg, \citealt{Kaiser1987, Percival2009}).
To date, nearly all velocity measurements have been obtained via
spectroscopy, mainly at optical wavelengths (\eg, \citealt{Abolfathi2018}).
One challenge to these spectroscopic measurements is the fundamental
degeneracy between the object's recessional velocity due to the expansion
of the universe and its peculiar velocity relative to that expansion.
The kinematic Sunyaev-Zel'dovich (SZ) effect signal, which is a Doppler
shift of cosmic microwave background (CMB) photons inverse Compton scattering
with a distribution of electrons, has long held
the promise of addressing this challenge by providing velocity
measurements relative to the fixed reference frame of the CMB 
(\citealt{Sunyaev1972, Sunyaev1980}; for a recent review see \citealt{Mroczkowski2018}).

However, measurements of the kinematic SZ effect signal have proven difficult,
mainly due to a lack of raw sensitivity but also due to contamination from a range
of unwanted astronomical signals (\eg, \citealt{Benson2003, Benson2004,
Kitayama2004, Zemcov2012}).
This situation is slowly changing, as a range of modern instruments have
been able to obtain tentative detections of the kinematic SZ effect in
resolved observations of exceptional individual galaxy clusters
with very high velocity sub-components \citep{Sayers2013, Adam2017} 
and in aggregate for large statistical samples 
\citep{Planck2016_XXXVII, Soergel2016, deBernardis2017}.
Looking forward, the next generation of instrumentation aims to advance
from these first detections to detailed studies using the
kinematic SZ effect (\eg, \citealt{Morandi2013, Mittal2018}).
While these SZ effect studies are unlikely to reach the velocity sensitivity
demonstrated by \hitomi\ in the central regions of nearby galaxy clusters,
they will ideally complement future X-ray observations from facilities
such as {\it XRISM}, {\it Athena}, and {\it Lynx} by providing
velocity measurements at higher redshifts and/or further from
the galaxy cluster's center.

\begin{deluxetable*}{cccccccc}
\tablecaption{Galaxy Cluster Sample}
\tablecolumns{8}
\tablewidth{0pt}
\tablehead{
\colhead{Name} &
\colhead{RA} &
\colhead{Dec} &
\colhead{Redshift} &
\colhead{\mfive} &
\colhead{140 GHz rms} &
\colhead{270 GHz rms} &
\colhead{Dynamical State} \\
\colhead{} &
\colhead{HH:MM:SS.s} &
\colhead{DD:MM:SS} &
\colhead{} &
\colhead{$10^{14}$ M$_{\sun}$} &
\colhead{\mjysr} &
\colhead{\mjysr} &
\colhead{}}
\startdata
\asix & 08:42:57.6 & $+$36:21:57 & 0.282 & $17.1 \pm 2.9$ & 0.010  & 0.025  (B) &  LOS-merger \\
\aeight & 14:01:01.9 & $+$02:52:40 & 0.253 & $12.3 \pm 1.4$ & 0.011  & 0.031  (B) &  relaxed \\
\moone & 00:18:33.4 & $+$16:26:13 & 0.546 & $16.5 \pm 2.5$ & 0.013  & 0.019  (A) &  LOS-merger \\
\motwo & 00:25:29.9 & $-$12:22:45 & 0.584 & \phn$7.6 \pm 0.9$ & 0.011  & 0.025  (A) &  POS-merger \\
\mfour & 04:54:11.4 & $-$03:00:51 & 0.538 & $11.5 \pm 1.5$ & 0.010  & 0.024  (A) &  POS-merger \\
\mseven & 07:17:32.1 & $+$37:45:21 & 0.546 & $24.9 \pm 2.7$ & 0.020  & 0.020  (B) &  LOS-merger \\
\mtwo & 21:29:25.7 & $-$07:41:31 & 0.589 & $10.6 \pm 1.4$ & 0.015  & 0.023  (A) &  LOS-merger \\
\roone & 01:52:41.1 & $-$13:58:07 & 0.833 & \phn$7.8 \pm 3.0$ & 0.014  & 0.014  (A) &  POS-merger \\
\rtwo & 12:26:57.9 & $+$33:32:49 & 0.888 & \phn$7.8 \pm 1.1$ & 0.015  & 0.021  (B) &  POS-merger \\
\rthree & 13:47:30.8 & $-$11:45:09 & 0.451 & $21.7 \pm 3.0$ & 0.013  & 0.032  (A) &  POS-merger \\
\enddata
\tablecomments{The ten galaxy clusters that were included in our study. The 
coordinates (corresponding to the X-ray centroid), redshifts, and masses were taken
from \citet{Sayers2013b}, and the masses were determined from \chandra\ data based on
the procedures described in \citet{Mantz2010}. The rms noise values are given for
1\arcmin\ pixels based on the average subtraction algorithm that was used for the SZ effect analysis
described in Section~\ref{sec:SZ_red}. 
Due to the presence of noise on large angular scales as a result of fluctuations in
atmospheric brightness, these values cannot be directly converted to
an rms in a different size pixel.
(A) denotes 270 GHz data from AzTEC and (B) denotes 270 GHz data
from Bolocam. See the text 
in Section~\ref{sec:sample} for a more detailed 
description of the dynamical state for each galaxy cluster.
}
\label{tab:sample}
\end{deluxetable*}

In this work, we used observations from Bolocam and AzTEC/ASTE,
along with ancillary data from \spire, \chandra, \planck, and 
the {\it Hubble Space Telescope} (\hst)
to obtain resolved images of the SZ effect signal towards
a sample of ten galaxy clusters. This analysis was built upon the previous work
of \citet{Sayers2013}, who used a subset of these data to detect
the kinematic SZ effect signal towards one of the galaxy clusters
in our sample, \mseven.
In Section~\ref{sec:sample}, we describe the sample of ten 
galaxy clusters in detail.
The datasets and their associated reduction 
(including the reconstruction of lens models) are then presented
in Sections~\ref{sec:data} and \ref{sec:reduction}.
Our fits to the SZ effect signals, and the galaxy cluster-averaged
bulk velocities obtained from these fits are given
in Section~\ref{sec:bulk}.
We then present resolved images of the SZ effect signals in
Section~\ref{sec:imaging}.
Finally, we provide a summary of our analysis in
Section~\ref{sec:summary}.

\section{Galaxy Cluster Sample} 
\label{sec:sample}

This study was focused on a sample of ten massive galaxy clusters with available
data from Bolocam/AzTEC, {\it Herschel}-SPIRE, \chandra, and \hst.
A brief description of the dynamical state of 
each galaxy cluster is given below, with a summary in
Table~\ref{tab:sample}.

\begin{enumerate}[align=left]
  \item[\asix:] \citet{Girardi2006}, based on \chandra\ X-ray and galaxy cluster member
    spectroscopic measurements, suggested that this system is undergoing a complex
    merger mainly along the line of sight (LOS). This complex merger scenario is 
    further supported by the detailed study of its giant radio halo by \citet{Macario2010}.
    \citet{Rossetti2013} also found indications for a merger mainly along the LOS.
  \item[\aeight:] This galaxy cluster was among the first targets of both \chandra\ and \xmm,
    and that imaging revealed a highly relaxed morphology \citep{Peterson2001, Schmidt2001}.
    A wide range of subsequent studies have supported the conclusion that this is one
    of the most relaxed known galaxy clusters (\eg, \citealt{Mantz2015}).
  \item[\moone:] \citet{Solovyeva2007} found this galaxy cluster to be undergoing a merger
    based on \chandra\ and \xmm\ data, and \citet{Piffaretti2003} found evidence for
    LOS elongation based on a joint X-ray and SZ effect analysis. \citet{Mann2012}, in their
    systematic study of 108 galaxy clusters to search for binary mergers, found this galaxy cluster
    to have a morphological code of 3 on their scale of 1--4, with 4 being the
    most likely to be undergoing a major merger. However, the reason it was not classified
    as a 4 was the relatively small offset between the BCG and the X-ray peak, which
    would be consistent with a merger primarily along the LOS.
  \item[\motwo:] This galaxy cluster is a dramatic plane of sky (POS) merger, similar to the Bullet
    Cluster, and has been studied in detail by several groups
    \citep{Bradac2008, Ma2010, Riseley2017, Cibirka2018}. \citet{Mann2012}
    listed this galaxy cluster as a textbook example of a binary merger and gave it a morphological
    code of 4.
  \item[\mfour:] Both \citet{Donahue2003} and \citet{Jeltema2005} found the
    X-ray morphology of this galaxy cluster to be elongated in the E--W direction
    in the POS, indicating a possible merger along that orientation.
    Furthermore, \citet{Mann2012} gave this galaxy cluster a morphological code of 3,
    and found a significant offset between the BCG and the X-ray peak.
  \item[\mseven:] The detailed analysis of \citet{Ma2009} showed this galaxy cluster
    to be a complex merger with a significant component along the LOS.
    In particular, they identified four merging subclusters in the system, and they
    labeled the largest subcluster, which is located slightly SE of the X-ray
    center, as ``C''. Approximately 1.5\arcmin\ NW of ``C'' is subcluster ``B'',
    which appears to be moving with a LOS velocity of $+3000$~\kms\ relative
    to ``C''.
    This scenario was further supported by a range of subsequent analyses, 
    including two based on kinematic SZ effect measurements
    \citep{Mann2012, Sayers2013, Adam2017, vanWeeren2017}.
  \item[\mtwo:] This galaxy cluster was given a morphological code of 3 by
    \citet{Mann2012}, and was described in that paper as a complex merger
    that is occurring primarily along the LOS.
  \item[\roone:] \citet{Maughan2006}, based on \xmm\ data, found that this galaxy cluster
    is undergoing a merger along two main axes, both oriented in the POS.
    A consistent merger scenario was found by \citet{Molnar2012} based on the offset
    between the X-ray and SZ effect signal peaks.

\begin{deluxetable*}{cccccc}
\tablecaption{Instrument Band Centers}
\tablecolumns{9}
\tablewidth{0pt}
\tablehead{
\colhead{Observing Band} & 
\colhead{Blackbody} &
\colhead{Thermal SZ} &
\colhead{Kinematic SZ} &
\colhead{Synchrotron} &
\colhead{Thermal Dust}}
\startdata
Bolocam 140 GHz & 140.5 GHz & 139.3 GHz & 140.1 GHz & 139.2 GHz & 141.2 GHz \\
Bolocam 270 GHz & 270.9 GHz & 274.9 GHz & 268.0 GHz & 267.7 GHz & 272.2 GHz \\
AzTEC 270 GHz   & 271.3 GHz & 275.5 GHz & 268.1 GHz & 267.8 GHz & 272.8 GHz \\
\enddata
\tablecomments{Effective instrument band centers for sources with various
SEDs. The overall spectral bandpass for each instrument
is a combination of the lab-measured spectral bandpass and 
the average atmospheric transmission at each site
computed from the ATM code described in 
\citet{Pardo2001a}, \citet{Pardo2001b}, and \citet{Pardo2005} (assuming 1.0 mm of precipitable water
vapor for AzTEC on the ASTE telescope and 1.5 mm of precipitable water
vapor for Bolocam on the CSO telescope).
From left to right the columns show the band center for a thermal blackbody
source in the Rayleigh-Jeans limit, the thermal SZ effect for a source with \te~$= 10$~keV,
the kinematic SZ effect for a source with \te~$= 10$~keV, a synchrotron source with a power
law exponent of $-0.7$, and a thermal dust source with an SED given
by Equation~\ref{eqn:sed} with $T_{\textrm{\tiny d}} = 15$~K.
}
\label{tab:bands}
\end{deluxetable*}

  \item[\rtwo:] \citet{Maughan2007} found evidence for merger activity in a joint
    \chandra\ and \xmm\ analysis. The weak lensing analysis of \citet{Jee2009} further supported
    a merger scenario. They found a large POS separation of the clumps, indicating that
    the merger may be oriented primarily along the POS. More recent SZ effect imaging from \citet{Korngut2011}
    and \citet{Adam2015} provided additional evidence for a POS merger scenario.
  \item[\rthree:] A range of independent analyses have found evidence for 
    a merger in the core region of this galaxy cluster, oriented along the SW--NE 
    direction and primarily in the POS \citep{Mason2010, Johnson2012, Plagge2013,
    Kreisch2016, Ueda2018}.
\end{enumerate}

\section{Datasets} 
\label{sec:data}

\subsection{Bolocam 140 GHz}

All of the galaxy clusters in our sample were imaged with Bolocam
at 140~GHz,\footnote{Throughout this work we refer to the SZ effect bands
as ``140 GHz'' and ``270 GHz''. The precise band centers for a range
of source spectra are given in Table~\ref{tab:bands}.} 
and all of those data have been used in previous analyses
({\it e.g.}, \citealt{Sayers2013, Czakon2015}) and are publicly 
available.\footnote{\url{https://irsa.ipac.caltech.edu/data/Planck/release_2/ancillary-data/bolocam/}}
The images have a point-spread function (PSF) with a solid angle that
corresponds to a Gaussian with a full-width at half maximum (FWHM) of
59.2\arcsec.
The data were collected in 10 minute observations using a sinusoidal
Lissajous pattern with differing periods in the right ascension and 
declination directions,
resulting in a coverage that drops to half its peak value at a radius of 
5--6\arcmin.
We obtained approximately 100 such individual observations per galaxy cluster.
Astrometry, with an rms uncertainty of $\simeq 5\arcsec$, was computed based on frequent
observations of nearby bright objects.

Nightly observations of Uranus and Neptune were used to calibrate the
detector response, and a single empirical fit as a function of atmospheric
opacity, accurate to 1.0\%, was
computed for all of the nights within a given observing run (typically
$\sim 10$ nights, see \citealt{Sayers2012}). 
For this work, we used the planetary models from
\citet{Griffin1993} rescaled based on the recent measurements from \planck,
which are accurate to 0.6\% at 140~GHz \citep{Planck2017_LII}. While our
empirical fit accounted for changes in band-averaged 
atmospheric transmission as a function of opacity, it did not account
for the slight changes in the spectral shape of the atmospheric transmission,
which we estimated to produce a 0.2\% rms uncertainty in our calibration
(see \citealt{Sayers2012}).

In addition, in transferring the calibration from the point-like planets to resolved SZ
effect surface brightness measurements there was an additional uncertainty due
to our characterization of the PSF solid angle, which we estimated to
be 1.2\% based on the quadrature sum of two separate uncertainties. 
First, \citet{Sayers2009} measured the per-detector solid
angle with an rms of 3.1\%, with no evidence for variation from
detector to detector. Therefore, averaging over the $\simeq 100$ optical
detectors resulted in a 0.3\% rms measurement uncertainty. 
Second, the measured solid angle was based on a source spectrum matching
that of Uranus and Neptune, which were used for the PSF calibration measurements.
We assumed the PSF was diffraction limited, which means its solid
angle was different for sources with different spectral shapes, such
as the thermal and kinematic SZ effect signals. To account for this difference
we included an additional rms uncertainty of 1.2\%, equal to the average
difference in diffraction-limited PSF solid angle for the effective band
centers of the thermal and kinematic SZ effect signals compared to the effective
band centers for Uranus and Neptune. 

In total, we estimated our calibration to be accurate to an rms uncertainty
of 1.7\%
(see Table~\ref{tab:cal}).

\begin{deluxetable*}{cccccccc}
\tablecaption{SZ Effect Calibration Uncertainty}
\tablecolumns{8}
\tablewidth{0pt}
\tablehead{
\colhead{Observing Band} & 
\colhead{Measurement} &
\colhead{\planck\ Abs} &
\colhead{Extrapolation} &
\colhead{PSF} &
\colhead{Atm Trans} &
\colhead{Total}}
\startdata
Bolocam 140 GHz & 1.0\% & 0.6\% & 0.0\% & 1.2\% & 0.2\% & 1.7\% \\
Bolocam 270 GHz & 1.0\% & 0.7\% & 1.3\% & 2.6\% & 0.3\% & 3.2\% \\
AzTEC 270 GHz   & 1.2\% & 0.7\% & 1.3\% & 2.8\% & 0.3\% & 3.4\% \\
\enddata
\tablecomments{Summary of the SZ effect calibration uncertainty. The columns show
the observing band, the uncertainty due to measurement error in the observations
of Uranus and Neptune, the absolute calibration uncertainty from \planck, 
uncertainties due to the extrapolation from the \planck\ observing bands
to our observing bands, measurement uncertainties on the PSF solid angle,
uncertainties due to changes in the shape of the atmospheric transmission 
spectrum as a function of opacity, and the total uncertainty.
}
\label{tab:cal}
\end{deluxetable*}

\subsection{Bolocam 270 GHz}

Four of the galaxy clusters in our sample were observed with Bolocam at 270~GHz,
using the same observing strategy detailed above for the 140~GHz data.
The 270~GHz Bolocam images have PSFs with a solid angle that corresponds 
to a Gaussian with a FWHM of 33.2\arcsec.
Compared to the 140~GHz data, 
some of the uncertainties on the calibration were slightly different
for the 270~GHz data
(see Table~\ref{tab:cal}).
Specifically, the absolute \planck\ measurements were accurate to 0.7\%
and the PSF solid angle characterization resulted in a 2.6\% calibration
uncertainty (0.6\% due to measurement uncertainty and 2.5\% due
to the differing effective band centers of the thermal and kinematic SZ effect signals).
In addition, unlike at 140~GHz, there was no \planck\ band centered near
our observing band at 270~GHz. As a result, we extrapolated the \planck\
measurements at 220 and 350~GHz, and we estimated this extrapolation 
resulted in a 1.3\% uncertainty based on the deviations obtained from calibrating
the \cite{Griffin1993} model at one of those frequencies and then comparing its
prediction to the measured value at the other frequency.
The total calibration uncertainty was determined to have an rms uncertainty of 3.2\%.

\subsection{AzTEC 270 GHz}

Six of the galaxy clusters in our sample were observed with AzTEC at 270~GHz
from the ASTE telescope (AzTEC was built as a nearly exact replica of Bolocam,
see \citealt{Wilson2008}). The scan pattern used for these observations
was very similar to the Lissajous
used in the Bolocam observations, and the resulting coverage was similar.
The PSF in the images has a solid angle that corresponds to a Gaussian
with a FWHM of 30.4\arcsec.
The calibration uncertainty was nearly identical to the 270~GHz Bolocam data,
although the slightly higher measurement uncertainty resulted in a total
calibration uncertainty with an rms of 3.4\% (see Table~\ref{tab:cal}).

\subsection{\spire}

All of the galaxy clusters in our sample were observed by \spire\ as part of 
either the {\it Herschel} Multi-tiered Extragalactic Survey 
(HerMES, \citealt{Oliver2012}) or the {\it Herschel} Lensing Survey
(HLS, \citealt{Egami2010}). \spire\ was a three-band photometric imager
operating at 600, 850, and 1200~GHz with PSFs with FWHMs of
18.1\arcsec, 25.2\arcsec, and 36.6\arcsec \citep{Griffin2010}.
The absolute calibration uncertainty of the \spire\ data was 5.5\%
for unresolved sources, and was verified by cross-calibrating with
\planck\ \citep{Bertincourt2016}.
In all cases, the \spire\ coverage was sufficient to produce images
in all three bands comparable in size to the Bolocam and AzTEC
images.

\subsection{\chandra}

Each galaxy cluster was observed in one or more \chandra\ X-ray imaging observations.  
The observation identification numbers (ObsIDs) and  exposure times are listed 
in Table~\ref{tab:xray_obs}.  Additionally, we provide information about whether 
the observation was taken with the imaging or spectroscopic Advanced CCD Imaging 
Spectrometer (ACIS-I or ACIS-S, respectively).  Since both instruments were used 
in imaging mode, this only impacted the sensitivity, background, and field of 
view of the exposure.  Since each CCD array subtends $8\arcmin \times 8\arcmin$, 
observations with either ACIS-I or ACIS-S covered a sufficiently large field 
of view for this analysis. 

\begin{deluxetable*}{cccc}
\tablecaption{{\it Chandra} X-ray Observations}
\tablecolumns{4}
\tablewidth{0pt}
\tablehead{
\colhead{Name} & 
\colhead{Inst.} &
\colhead{ObsIDs} &
\colhead{Usable Exp.\ Times (ksec)}
}
\startdata
\asix   & ACIS-I & 4217 & 19.2 \\
\aeight   & all ACIS-I & 6880,6881,7370 & 115.9,36.3,39.5 \\
\moone & ACIS-I & 520 & 64.1 \\
\motwo & all ACIS-I & 3251,5010,10413,10786,10797 & 18.0,23.8,75.6,13.7,23.8 \\
\mfour & ACIS-I,ACIS-S & 529,902 & 13.7,41.9 \\
\mseven & all ACIS-I & 1655*,4200,16235,16305 & ---,54.9,67.3,89.9\\
\mtwo & all ACIS-I & 3199*,3595 & ---,18.2\\
\roone   & ACIS-I & 913 & 34.7 \\
\rtwo   & all ACIS-I & 3180,5014 & 29.1,30.8 \\
\rthree & all ACIS-I & 3592,13516,13999,14407 & 56.6,39.0,54.4,63.0\\
\enddata
\tablecomments{Summary of the \chandra\ ACIS-S and ACIS-I imaging 
exposures used for X-ray spectroscopic temperature analysis.
Exposure times reported indicate the usable time on source after flare filtering.
ObsID 3199 was excluded from the spectroscopic analysis due to flare contamination.
ObsID 1655 was excluded due to the relative brevity of the observation and potential calibration differences.}
\label{tab:xray_obs}
\end{deluxetable*}

\subsection{\hst}
\label{sec:hst}

We reconstructed lens models for the ten galaxy clusters of our sample using
multiband \hst\ imaging, essential for the identification of multiple-image
constraints. Although the lens models were largely based on existing models,
for completeness we describe the latest
available \hst\ imaging which enabled these models.
Eight galaxy clusters from our sample were imaged extensively with both optical
and near-infrared broadbands in the framework of large lensing surveys such
as Cluster Lensing And Supernova with \hst\ (CLASH; PI: Postman; \mseven, \mtwo,
\rthree, \rtwo, Reionization Cluster Survey (RELICS; PI:
Coe; \asix, \moone, \motwo, \roone), and the \hst\ Frontier Fields
(PIs: Mountain, Lotz; \mseven). For the remaining two galaxy clusters, reduced
images were downloaded from the \hst\ Legacy Archive, taken in program ID
11591 for both \aeight\ and \mfour\ (PI: Kneib), and programs IDs
10493 (PI: Gal-Yam), 9722 (PI: Ebeling), 9292 (PI: Ford), and 9836 (PI:
Ellis), for \mfour. The typical depth for most galaxy clusters was
$\sim26.5-27$ AB per band, and the typical pixel
scale was 0.05\arcsec--0.06\arcsec\ per pixel. 
Details of the lens modeling are given
in Section~\ref{sec:hst_red}.

\section{Data Reduction} 
\label{sec:reduction}

\subsection{Bolocam and AzTEC}
\label{sec:SZ_red}

The Bolocam data at 140 and 270~GHz, along with the AzTEC data
at 270~GHz, were reduced in a uniform manner
using the analysis pipeline described in detail in \citet{Sayers2011}.
For the SZ effect analysis, a template of the atmospheric brightness fluctuations was 
computed by averaging the signal from all of the detectors at each
time sample within a single $\simeq 10$ minute observation.
A single correlation coefficient between each detector's data
stream and the template was then computed, and the template was 
subtracted after rescaling by this correlation coefficient.
For Bolocam, the correlation coefficient was computed using only
the data within a narrow bandwidth of the two fundamental
Lissajous scan frequencies.
For AzTEC, where the scan frequencies were constantly modulated,
we instead computed the correlation coefficient using all of the
data within the bandwidth 0.5--2.0~Hz.
After this subtraction, a high-pass filter was applied to the
data streams, with a characteristic frequency of 250~mHz
for the 140~GHz data and 500~mHz for the 270~GHz data.
The template removal and high-pass filter resulted in a non-unity
transfer function for astronomical signals, and we computed
a single transfer function for the two-dimensional image
of each galaxy cluster at each observing frequency according to
the procedure described in \citet{Sayers2011}.

At 270~GHz, for both AzTEC and Bolocam, we also performed a second
data reduction using an adaptive principal component analysis (PCA)
in place of the average template subtraction 
\citep{Laurent2005, Aguirre2011}. The adaptive PCA method
was not as effective
as the average template subtraction for recovering the SZ effect signal from 
the galaxy cluster, but it was better for detecting unresolved objects
\citep{Sayers2013}.

\begin{deluxetable*}{ccccccc}
\tablecaption{DSFG Detections}
\tablecolumns{7}
\tablewidth{0pt}
\tablehead{
\colhead{Name} & 
\colhead{270 GHz Det.} &
\colhead{270 GHz Lim.} &
\colhead{\spire\ Det.} &
\colhead{600 GHz Lim.} &
\colhead{Counterparts}}
\startdata
\asix & \phn4 & 4.00 mJy & 121 & \phn4.64 mJy & \phn3 \\
\aeight & \phn2 & 4.84 mJy & \phn57 & \phn9.16 mJy & \phn1 \\
\moone & 23 & 2.60 mJy & 111 & \phn4.42 mJy & 19 \\
\motwo & 11 & 3.20 mJy & 123 & \phn4.68 mJy & \phn9 \\
\mfour & 19 & 2.96 mJy & \phn38 & \phn8.62 mJy & \phn9 \\
\mseven & 13 & 3.76 mJy & 110 & \phn5.58 mJy & 13 \\
\mtwo & 20 & 3.24 mJy & \phn12 & 11.98 mJy & \phn8 \\
\roone & 22 & 2.20 mJy & \phn60 & \phn6.62 mJy & 16 \\
\rtwo & \phn5 & 4.52 mJy & \phn27 & \phn10.00 mJy & \phn4 \\
\rthree & \phn9 & 3.60 mJy & \phn88 & \phn4.52 mJy & \phn9 \\
\enddata
\tablecomments{Summary of the detected point-like sources presumed to be
DSFGs. The columns give the number of sources detected in the 270~GHz
image, the 270~GHz detection limit at $\textrm{S/N} = 4$, the number
of sources detected by \spire, the 600~GHz detection limit at 
$\textrm{S/N} = 2$ (without accounting for noise from source confusion),
and the number of 270~GHz detections with a counterpart
identified in the \spire\ detections.
}
\label{tab:smgs}
\end{deluxetable*}

Regardless of the subtraction algorithm, the noise properties of the 
images were estimated using a set of 1000 random realizations 
based on the procedure given in \citet{Sayers2016b}.
First, 1000 jackknife realizations were generated by creating images
after randomly selecting half of the individual observations and
multiplying their data by $-1$. On average, this procedure 
removed all of the astronomical
signals while preserving the noise properties of the instrument
and the atmospheric fluctuations. To each of these 1000 jackknife images,
a random realization of the primary CMB fluctuations, the 
background population of dusty star-forming galaxies (DSFGs)
that comprise the cosmic infrared background (CIB), 
and the population of radio galaxies were added.
Each instrument's PSF and subtraction-dependent signal transfer function
was accounted for prior to adding these astronomical source 
realizations.
In order to fully capture any correlations in these unwanted astronomical
signals between 140 and 270~GHz, we did not generate separate
realizations at the two observing frequencies.
Instead, a single realization was scaled to both frequencies.

After producing the images, along with their associated
noise realizations, we then jointly 
fitted an elliptical generalized NFW (gNFW) model
\citep{Nagai2007}
to the 140~GHz Bolocam images and the \planck\ all-sky \ymap\
\citep{Planck2016_XXII},
according to the method detailed in \citet{Sayers2016} which
fully accounted for the Bolocam transfer function and the \planck\
and Bolocam PSFs.
For these fits, the normalization and scale radius of the model
were varied while fixing the three power law exponents $\alpha$, $\beta$, and $\gamma$
to the best fit values of \citet{Arnaud2010}.
For the radial scales typically probed by our data, 
$\simeq 0.3$\radius--3.0\radius,\footnote{
The values of \radius\ used in this work
were taken from \citealt{Czakon2015}, where they were
computed from \chandra\ X-ray data using a scaling relation 
between gas mass and total mass.}
this model had sufficient freedom to provide a good fit quality
(see \citealt{Sayers2011} and \citealt{Czakon2015}), particularly since
ellipticity in the POS was allowed.
Furthermore, while a range of more recent observational studies have found different
best-fit values of $\alpha$, $\beta$, and $\gamma$ (\eg, \citealt{Sayers2013, 
Planck2013_V, Ghirardini2018}), the actual profile shapes are 
in excellent agreement owing to the strong degeneracies between the parameters,
particularly when the scale radius is allowed to vary, as it was in our analysis.
We therefore do not expect any significant biases due to our choice of
model to describe the shape of the SZ effect signal.

The resulting best-fit gNFW model was then subtracted from the adaptive-PCA-reduced
270~GHz images, accounting for the transfer function and PSF
of those images. This subtraction removed most of the SZ effect signal,
leaving the background CIB as the dominant astronomical signal in the images.
We then used \textsc{StarFinder} to detect all of the 
unresolved objects with a $\textrm{S/N} > 4$ from the resulting
images. We typically detected $\simeq 10$ such objects in each image, all
of which were presumed to be DSFGs (see Table~\ref{tab:smgs}).
As detailed below in Section~\ref{sec:spire_red}, 
\spire\ was more sensitive to the signal from DSFGs, and typically
detected an order of magnitude more objects.

For the next step in our analysis, we returned to the 140 and 270~GHz SZ 
effect images created
using the average template subtraction. From these images, we subtracted
all of the radio galaxies listed in \citet{Sayers2013c} and all of 
the DSFGs detected in the 270~GHz images and/or the \spire\ images.
To subtract the radio galaxies, the power law fits from \citet{Sayers2013c}
were extrapolated to 140 and 270~GHz. 
The DSFGs were categorized into three groups, with a slightly different
procedure used to subtract the sources from within each of these groups.
The first group included DSFGs detected at 270~GHz without a counterpart
identified in the \spire\ detections.
These were subtracted from the 270~GHz data based on their detected flux
density, and from the 140~GHz data based on a rescaling of the flux
density according to $\nu^{2.5}$.
The second group included DSFGs detected at 270~GHz which had a \spire\
counterpart.
For these sources, the 270~GHz and \spire\ three-band measurements
were simultaneously fitted to a greybody SED of the form:
\begin{equation}
  \label{eqn:sed}
  F(\nu) = F_0 (1 - e^{(-\nu/\nu_0)^{\beta}}) B(\nu, T_{\textrm{\footnotesize d}})
\end{equation}
where the values of the normalization $F_0$ and the dust temperature
$T_{\textrm{\footnotesize d}}$ were varied, $\nu$ is the observed frequency, 
$\nu_0 = 3000$~GHz \citep{Draine2006}, 
$\beta = 1.95$ is the dust emissivity spectral index,\footnote{
Given the noise and spectral coverage of the data, we were unable to
robustly constrain the values of both $\beta$ and $T_{\textrm{\tiny d}}$
for a single source. We therefore fixed the value of $\beta$
in our fits. To determine what value of $\beta$ to use,
we compared the measured flux density in 
the 270~GHz images at the position of every \spire\ detection
to the flux density computed from a greybody fit solely to the three-band
\spire\ images extrapolated to 270~GHz for a range of fixed $\beta$ values. 
On average, the two flux densities 
agree within the measurement noise for $\beta = 1.95 \pm 0.11$.
This was consistent with the value of $\beta$ found in several other
recent studies (\eg, \citealt{Magnelli2012, Smith2013}).}
and 
$B(\nu, T_{\textrm{\footnotesize d}})$ is the Planck function.
The sources were then subtracted from the 140 and 270~GHz images
based on the flux densities obtained from this greybody fit.
The third and final group included DSFGs detected by \spire\ that
were not associated with a 270~GHz detection.
These sources were subtracted in an analogous way to those 
in the second group,
except that the greybody SED was fit solely to the three-band \spire\
images.

This procedure for characterizing and subtracting the DSFGs was nearly
identical to what was described in detail in the appendix of
\citet{Sayers2013}. Since the quality
of the data used in this work was nearly identical to the data used
by \citet{Sayers2013}, the same overall implications were also true
and are summarized here.
In particular, all sources brighter than $\simeq 1$~mJy at 270~GHz were detected,
and some sources were detected down to a limit of $\simeq 0.1$~mJy at
270~GHz. In aggregate, these detected sources represent $\simeq 30$\%
of the total emission from the CIB at that frequency. As noted above,
most of the detections were made by \spire, and the AzTEC/Bolocam
detections typically amounted to only $\simeq 5$--10\% of the total CIB.
Even after subtracting $\simeq 30$\% of the CIB emission, the 
fluctuations due to the remaining sources added an rms per beam of
approximately 0.5~mJy, and these fluctuations
degraded our SZ effect constraints at 270~GHz by $\simeq 10$--20\%
compared to what would have been possible with perfect removal of the CIB.
While these undetected CIB sources added a non-negligible amount of noise, they
did not produce a measurable bias in the SZ effect constraints, likely because
their distribution was well described by a Gaussian rms given the PSF
size and noise level typical of our 270~GHz data.

In addition, we subtracted an image of the average apparent signal deficit in the CIB produced
by galaxy cluster lensing of the DSFG population when the brightest individual sources were
removed. This effect was first detected by \spire, and was used to estimate
the total brightness of the CIB \citep{Zemcov2013}.
In addition, \citet{Lindner2015} measured a lower than expected SZ effect
signal in \spire, and they speculated that this was due to lensing
of the CIB based on the previous \citet{Zemcov2013} results.
To estimate the lensing-induced CIB deficit in our SZ effect images,
we propagated a random realization of the CIB through
the lensing model determined from the \hst\ data (see Section~\ref{sec:hst_red}). 

\begin{figure*}
  \centering
  \includegraphics[width=\textwidth]{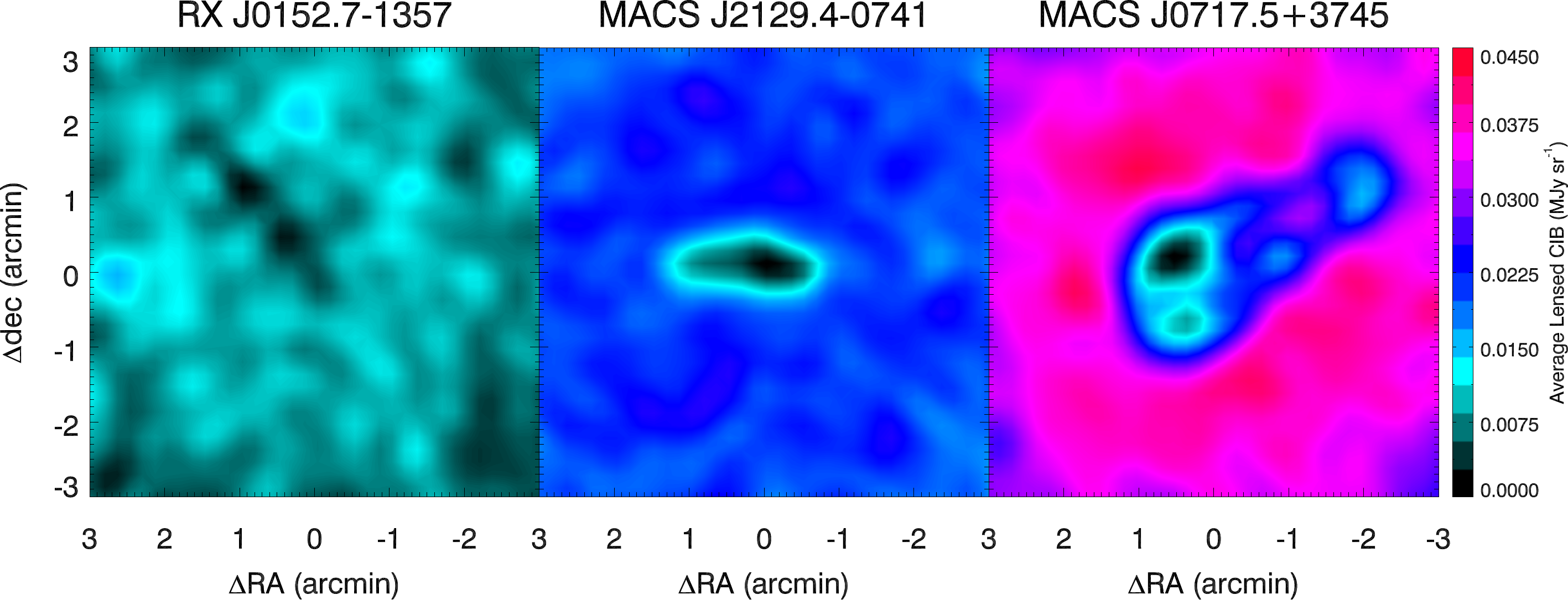}
  \caption{The average surface brightness of the background CIB for
  three of the galaxy clusters in our sample. In all cases individual bright sources
  were removed according to the procedure detailed in Section~\ref{sec:SZ_red},
  and spatial filtering according to the transfer function for the
  average template subtraction used
  for the SZ effect images has been applied. This filtering removed the mean signal level,
  and so all three images have been set to have a minimum signal of 0,
  and all are shown with the same color scale.
  On average, there was a deficit of brightness
  near the galaxy cluster center due to the combined effects of gravitational lensing
  and the subtraction of bright sources.
  From left to right, \roone\ was the weakest lens in our sample, \mtwo\
  was typical of our sample, and MACS \mseven\ was the strongest lens in our sample.}
  \label{fig:lensing}
\end{figure*}

\begin{deluxetable*}{cccc}
\tablecaption{The Impact of CIB Lensing on the Measured 270 GHz SZ Effect Brightness}
\tablecolumns{4}
\tablewidth{0pt}
\tablehead{
\colhead{Name} & 
\colhead{No Lensing Correction} &
\colhead{With Lensing Correction} &
\colhead{Difference}}
\startdata
\asix & 0.037 \mjysr & 0.046 \mjysr & 0.009 \mjysr \\
\aeight & 0.069 \mjysr & 0.080 \mjysr & 0.011 \mjysr \\
\moone & 0.097 \mjysr & 0.117 \mjysr & 0.020 \mjysr \\
\motwo & 0.043 \mjysr & 0.052 \mjysr & 0.009 \mjysr \\
\mfour & 0.081 \mjysr & 0.091 \mjysr & 0.010 \mjysr \\
\mseven & 0.109 \mjysr & 0.131 \mjysr & 0.022 \mjysr \\
\mtwo & 0.040 \mjysr & 0.053 \mjysr & 0.013 \mjysr \\
\roone & 0.081 \mjysr & 0.086 \mjysr & 0.005 \mjysr \\
\rtwo & 0.105 \mjysr & 0.124 \mjysr & 0.019 \mjysr \\
\rthree & 0.078 \mjysr & 0.087 \mjysr & 0.009 \mjysr \\
\enddata
\tablecomments{The 270 GHz SZ effect brightness towards each galaxy cluster
before and after accounting for the CIB deficit due to
gravitational lensing and the subtraction of bright sources. 
On average, the two values differed
by 0.013 \mjysr, or $\simeq 15$\% of the SZ effect brightness.
}
\label{tab:lensing}
\end{deluxetable*}

For each galaxy cluster we generated 100 such realizations. Individual
bright sources were then removed from these realizations in a way
that mimicked the procedure applied to the actual data, which resulted
in the detection limits given in Table~\ref{tab:smgs}. 
After removing the bright sources, each realization was then spatially
filtered based on the transfer function for the
data reduction using an average template subtraction.
The 100 realizations were then averaged for each galaxy cluster, and the result
was subtracted from the actual SZ effect images. 
While lensing can produce large brightness variations 
in the CIB due to the (rare) high magnification of intrinsicly bright DSFGs,
all such extremely bright objects were subtracted from both our
real data and the 100 lensed realizations.
As a result, the typical brightness fluctuations between the 100 lensed realizations
were well described by the unlensed CIB realizations already
included in our noise model.
Examples of the average
lensed CIB are shown in Figure~\ref{fig:lensing}. Based on the bulk
SZ effect fits described in 
Section~\ref{sec:bulk}, the typical deficit in the CIB due to
lensing was $\simeq 15$\% of the SZ effect brightness at 270~GHz
(see Table~\ref{tab:lensing}).

\subsection{\spire}
\label{sec:spire_red}

The three-band \spire\ images were used to search for and characterize
DSFG candidates. The data were reduced using the {\it Herschel} Interactive
Processing Environment (HIPE, \citealt{Ott2006, Ott2010}) and the
HerMES SMAP package \citep{Levenson2010, Viero2013}.
A list of DSFG candidates was compiled based on the SCAT procedure \citep{Smith2012},
with the requirement that each source have a $\textrm{S/N} > 2$ at both
600 and 850~GHz.
We found that many of the brighter DSFGs at 270~GHz were not detected
at 1200~GHz by \spire, and so we did not impose a S/N threshold on those
data.
This typically resulted in $\simeq 100$ DSFG candidates per galaxy cluster
(see Table~\ref{tab:smgs}).

\subsection{\chandra}

The \chandra\ data reduction and analysis closely followed the methods presented in 
\citet{Ogrean2015} and \citet{vanWeeren2017}, based on the publicly 
available scripts used in those previous analyses.\footnote{See 
\url{https://github.com/gogrean/MACS-J0717-Filament/blob/master/code/notebooks/}.} 
Briefly, the data were reprocessed to apply the latest calibration at the time, 
in this case {\tt CIAO 4.10} with {\tt CALDB 4.7.8}. Both of these tools were released 
sufficiently after each observation used for analysis that the calibration was 
stable/unchanging for newer releases.
In the case of observations taken in VFAINT mode, the {\tt check\_vf\_phaevents} 
option was used to provide additional filtering for background events.
As in \citet{Ogrean2015}, we extracted light curves from detector regions excluding 
point sources identified using {\tt wavdetect} as well as the galaxy cluster itself, 
and we used the {\tt CIAO} tool {\tt deflare} to identify periods of flaring. 
The resulting useful time on source, known as the ``good time interval'' (GTI), 
is reported in Table~\ref{tab:xray_obs}. We then extracted new events files using 
those GTIs, and those clean event files were used for all further X-ray analysis.

For the X-ray spectral analyses used to produce \te\ maps, the stowed ACIS background 
files were rescaled to match the high energy (10--12~keV) count rates off source 
(again, excluding regions with point source and galaxy cluster emission). These rescaled 
backgrounds were used as backgrounds in the spectral analysis.
The regions used for spectroscopy were selected using the contour binning method of 
\citet{Sanders2006}.\footnote{Known as {\tt contbin}, \url{https://github.com/jeremysanders/contbin}.} 
The parameters were chosen to ensure each region had sufficient counts (typically 
$\gtrsim3000$ background-subtracted counts from the inner portion of the galaxy cluster, 
though \mtwo\ had $\sim1800$) per spectral bin for reliable spectroscopy.
The spectral analysis was carried out jointly for all available datasets in {\it Sherpa} 
\citep{Freeman2001}, using the {\tt xsmekal} implementation of the Mewe-Kaastra-Liedahl (MeKaL) model.
The hydrogen column density $N_H$ was fixed to the value found using the {\tt CIAO} tool 
{\tt prop colden} to obtain an interpolation of the \citet{Dickey1990} value at the galaxy cluster location.
The redshift and abundance were also fixed in the analysis. 
In the case of abundance, several fits with abundance left free were also tested, and found 
not to differ significantly from fixing it to $Z=0.3~Z_\odot$.

\subsection{\hst}
\label{sec:hst_red}

\begin{figure*}
  \centering
  \includegraphics[width=\textwidth]{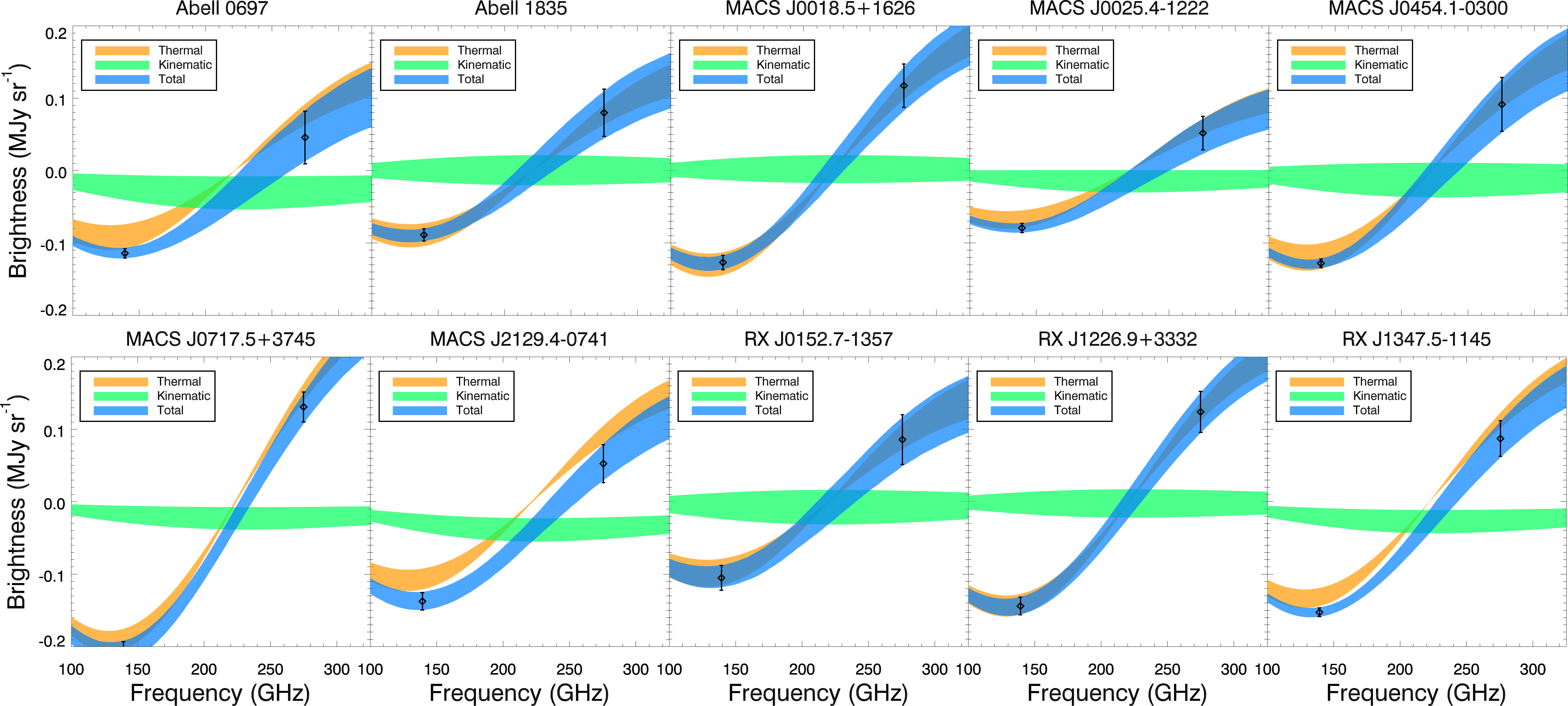}
  \caption{SZ effect spectral fits. The measured brightness within \radius\
  is given by the black points with error bars, and the 68\% confidence
  regions for the thermal,
  kinematic, and total SZ effect signals constrained by the data are
  shown in orange, green, and blue, respectively.}
  \label{fig:SZ_spec}
\end{figure*}

The \hst\ images used to construct the lens models
were already reduced, typically using standard procedures (most notably
multidrizzle, see \citealt{Koekemoer2011}).
For all of the galaxy clusters,
previous lensing analyses exist, including multiple image constraints. The
galaxy clusters were modeled here using parametrized forms, namely, double pseudo
isothermal elliptical mass distributions for the galaxy cluster galaxies following
common scaling relations, and elliptical NFW haloes for the galaxy cluster dark
matter clumps. For the CLASH galaxy clusters, we adopted
the \citet{Zitrin2015} ``PIEMDeNFW" mass models. 
For the \hst\ Frontier Fields galaxy cluster \mseven\
we remade and updated the model that is available on the \hst\ Frontier
Fields website.\footnote{\url{https://frontierfields.org}} 
For modeling the RELICS galaxy clusters, we adopted the constraints
from \citet{Cibirka2018} and \citet{Acebron2018}
and we constructed a model for \moone\ 
based on the constraints identified by
\citet{Zitrin2011}.
For the remaining two galaxy clusters, we constructed models based on the
multiple-image constraints listed in \citet{Richard2010}
and \citet{Zitrin2011}.
Then, as our aim here was to supply maps to lens the CIB
at radii well beyond the strong-lensing regime, and since our models were
constructed from analytic, parametric forms, we then regenerated the
strong-lensing models using the best-fit parameters from the above, but
covering a larger field of view extending to the weak lensing regime.
It should therefore be noted that these models have been
extrapolated, as they were only constrained using data from within the \hst\
field of view (solely strong lensing constraints, except for the CLASH galaxy clusters
where \hst\ weak lensing constraints were also used, see \citealt{Zitrin2015}).
We regenerated all of the models onto a $16\arcmin \times 16\arcmin$ map,
adopting a resolution of 0.25\arcsec\ per pixel. Using these
extended lens models we ray-traced different realizations of the background
DSFGs that comprise the CIB, as detailed in Section~\ref{sec:SZ_red}.

\section{Bulk Galaxy Cluster Velocities}
\label{sec:bulk}

\subsection{Method and Results}

Using the images produced in Section~\ref{sec:SZ_red}, from which radio galaxies, DSFGs,
and the average lensing-induced CIB signal deficit were subtracted, we fitted a parametric
model of the SZ effect signal to the data.
First, an elliptical gNFW model, with power law exponents $\alpha$, $\beta$, and $\gamma$
fixed to the values found by \cite{Arnaud2010}, was simultaneously fitted to the 140 GHz, 
270~GHz, and \planck\ \ymap\ data assuming a purely thermal SZ effect spectrum (i.e., zero 
kinematic SZ effect signal).
As in Section~\ref{sec:SZ_red}, the image transfer functions and PSFs were fully
accounted for in this fit.
After this initial fit, which was used to determine the two-dimensional shape of the 
SZ effect signal,
we then performed additional fits, separately to the 140~GHz and 270~GHz data,
where only the normalization of the gNFW model was allowed to vary.
This normalization was expressed in terms of the average surface brightness,
in \mjysr, within an aperture centered on the galaxy cluster and extending to a radius
of \radius.

After adding the best-fit SZ effect model to each of the 1000 noise realizations for each galaxy cluster,
an analogous two-step fit was performed, and the spread of normalization values obtained
from these 1000 fits was used to estimate the uncertainty on that parameter.
In addition, the $\chi^2$ values obtained from these 1000 fits were used to empirically
determine the fit quality based on a probability to exceed (PTE) using the 
procedure described in \citet{Sayers2011}.
The average PTE for the 10 clusters was 0.37, and only one cluster had a PTE
below 0.19 (\rtwo, which had a PTE of 0.01).
Therefore, even though many of these clusters are complicated mergers, the
elliptical gNFW model was sufficient to describe our data given their
noise and angular resolution. Furthermore, unlike X-ray observations,
which are proportional to ICM density squared, the SZ effect data are
linearly proportional to the ICM parameters (pressure for the thermal 
SZ effect and LOS velocity weighted by electron number density 
for the kinematic SZ effect).
As a result, merger-induced ICM sub-structures, which can significantly bias
similar bulk fits of smooth models to X-ray data, 
are much less problematic for fits to SZ effect data 
(\eg, \citealt{Motl2005, Kay2012}).

\begin{figure*}
  \centering
  \includegraphics[width=\textwidth]{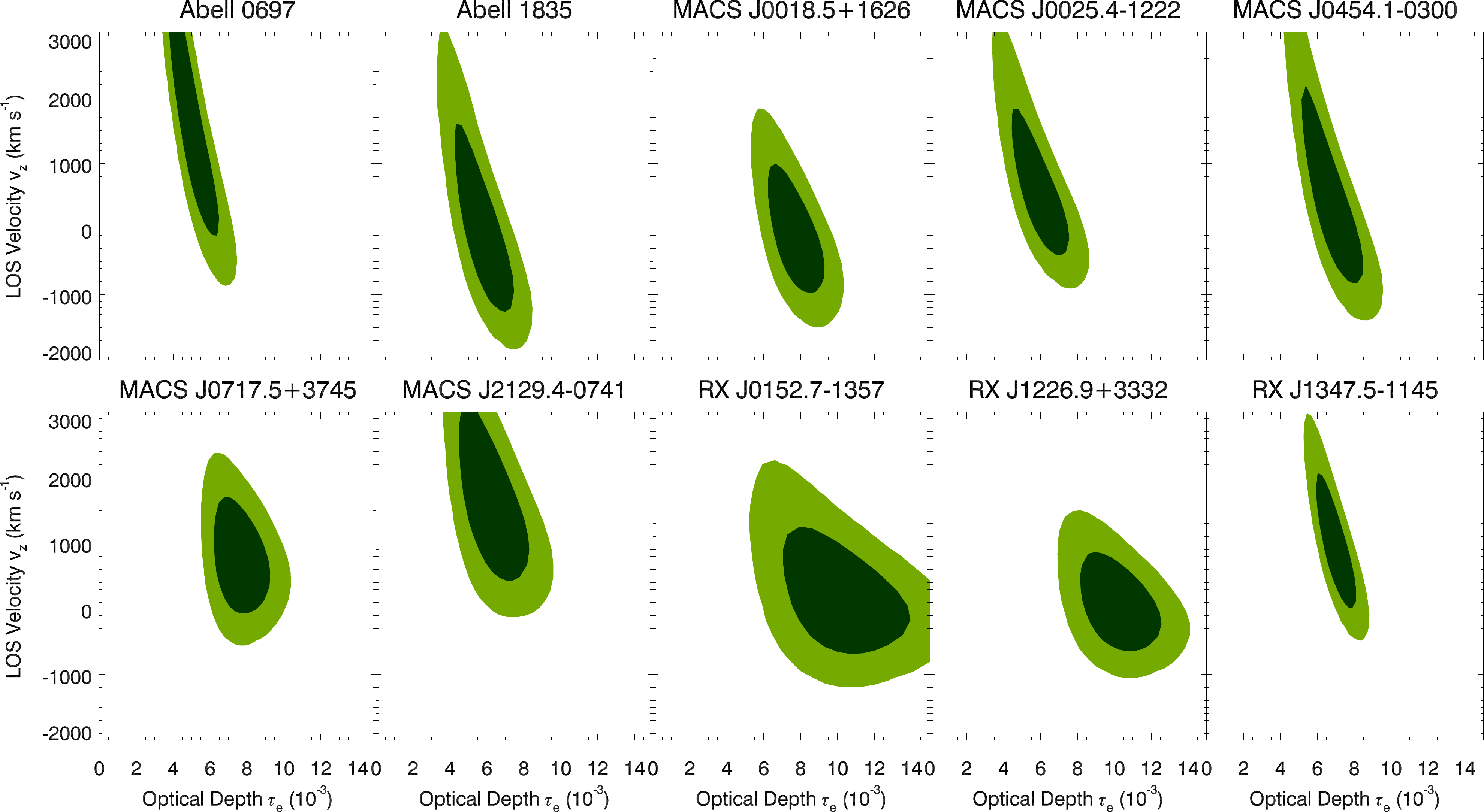}
  \caption{Constraints on the average electron optical depth \taue\ and
    LOS bulk velocity \vlos\ within \radius\ for each galaxy cluster. The dark green
    region encloses the 68\% confidence interval and the light green
    region encloses the 95\% confidence interval. The uncertainty
    on \vlos\ scales approximately like $1/$\taue, resulting in 
    a slightly curved degeneracy.}
  \label{fig:tau_vpec_contour}
\end{figure*}

The SZ effect brightness values obtained from the above procedure 
were then used to constrain the overall
bulk velocity of each galaxy cluster via the kinematic SZ effect.
Specifically, we assumed that each galaxy cluster was moving with a single
bulk LOS velocity and that its ICM was isothermal, with an electron temperature \te\
equal to the spectroscopic X-ray temperature measured by \chandra\
within \radius. 
Given the assumption of an isothermal ICM with \te\ measured by 
\chandra, the total brightness
from the thermal and kinematic SZ effect signals could be completely
specified in terms of the electron optical depth \taue\
and the bulk LOS velocity \vlos.
We used the \textsc{SZpack} software described in
\citet{Chluba2012, Chluba2013} to compute the SZ effect brightness for
a given set of parameters, including relativistic corrections and assuming
the effective thermal and kinematic SZ effect band centers given in 
Table~\ref{tab:bands}.
The results of these fits are shown in Figure~\ref{fig:tau_vpec_contour} and summarized
in Table~\ref{tab:results}.

\begin{deluxetable*}{ccccc}
\tablecaption{Derived ICM Parameters}
\tablecolumns{5}
\tablewidth{0pt}
\tablehead{
\colhead{Name} & 
\colhead{Temperature (keV)} &
\colhead{Optical Depth ($10^{-3}$)} &
\colhead{Bulk \vlos\ (\kms)} &
\colhead{Internal \vlos\ rms (\kms)}}
\startdata
\asix & \phn$8.99^{+0.53}_{-0.42}$ & \phn$4.88^{+0.86}_{-0.99}$ & $+1620^{+1250}_{-1500}$ & $1820^{+940}_{-940}$ \\
\aeight & \phn$7.66^{+0.13}_{-0.13}$ & \phn$5.69^{+1.04}_{-1.05}$ & \phn\phn$-70^{+\phn850}_{-\phn960}$ & $\le 1970$ (95\% CL) \\
\moone & \phn$8.30^{+0.49}_{-0.40}$ & \phn$7.64^{+0.99}_{-1.01}$ & \phn$-110^{+\phn610}_{-\phn640}$ & $\phn 810^{+600}_{-490}$ \\
\motwo & \phn$5.67^{+0.25}_{-0.24}$ & \phn$5.83^{+1.05}_{-1.04}$ & \phn$+530^{+\phn700}_{-\phn750}$ & $\le 1570$ (95\% CL) \\
\mfour & \phn$8.83^{+0.56}_{-0.50}$ & \phn$6.64^{+1.10}_{-1.10}$ & \phn$+370^{+\phn910}_{-1000}$ & $\le 1590$ (95\% CL) \\
\mseven & $12.83^{+1.42}_{-1.42}$ & \phn$7.55^{+0.92}_{-1.00}$ & \phn$+740^{+\phn530}_{-\phn560}$ & $1260^{+430}_{-360}$ \\
\mtwo & \phn$8.52^{+1.44}_{-1.14}$ & \phn$6.13^{+1.16}_{-1.23}$ & $+1570^{+\phn810}_{-\phn870}$ & $1170^{+560}_{-510}$ \\
\roone & \phn$4.72^{+0.56}_{-0.59}$ & \phn$9.96^{+2.12}_{-2.24}$ & \phn$+150^{+\phn560}_{-\phn610}$ & $\le \phn 960$ (95\% CL) \\
\rtwo & \phn$6.84^{+0.75}_{-0.59}$ & $10.13^{+1.47}_{-1.36}$ & \phn\phn$+40^{+\phn450}_{-\phn480}$ & $\le 1260$ (95\% CL) \\
\rthree & \phn$9.47^{+0.37}_{-0.29}$ & \phn$6.94^{+0.70}_{-0.71}$ & \phn$+950^{+\phn640}_{-\phn680}$ & $\le \phn 860$ (95\% CL) \\\enddata
\tablecomments{Best-fit values and 68\% confidence intervals for the
ICM parameters derived in our analysis. The temperature constraints 
were obtained from \chandra\ (not including the assumed 10\% systematic uncertainty), 
and the optical depth and bulk velocity
constraints were obtained from our SZ effect fits. The internal
\vlos\ rms constraints were obtained from the resolved SZ effect maps
within \radius. The 68\% confidence interval for six clusters
is consistent with an internal \vlos\ rms of zero, and 95\%
confidence level upper limits are given for these clusters.
}
\label{tab:results}
\end{deluxetable*}

\begin{figure}
  \centering
  \includegraphics[width=0.4\textwidth]{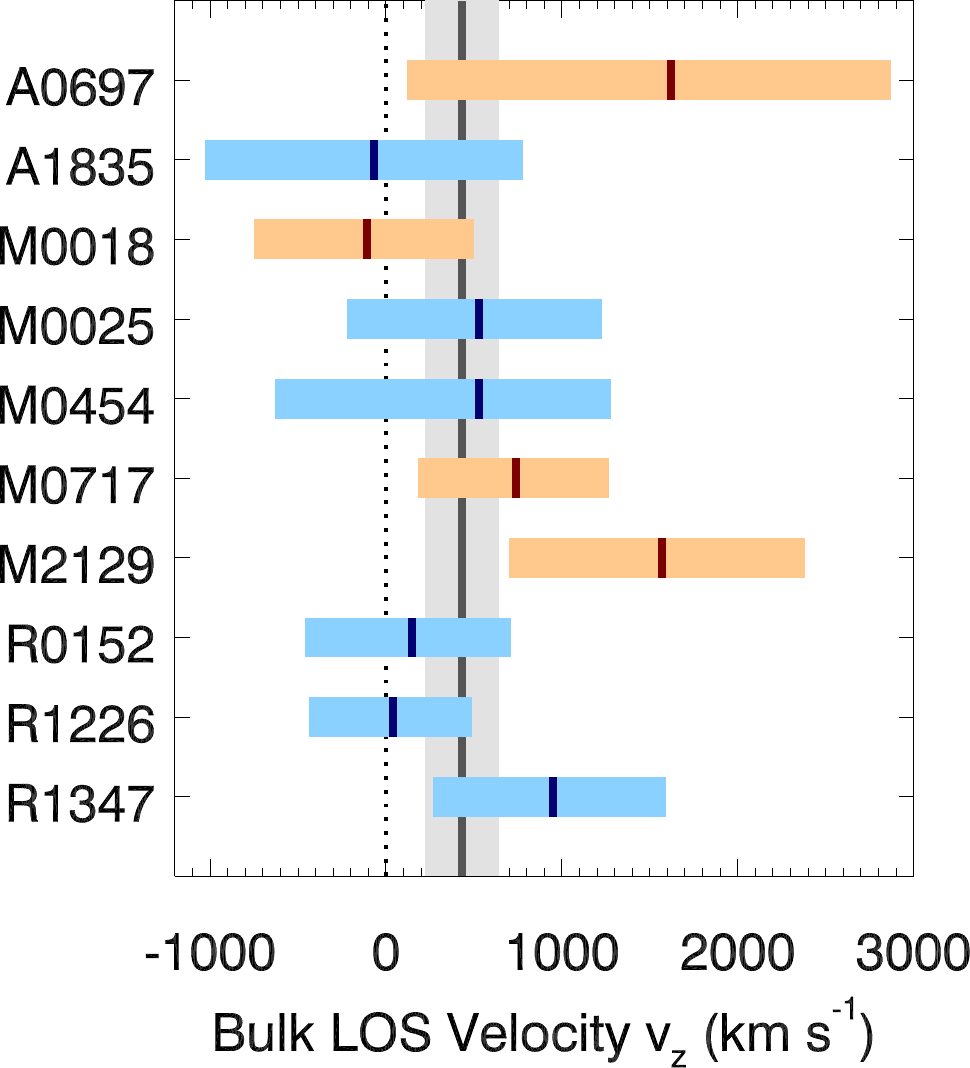}
  \caption{The best-fit bulk LOS velocity \vlos\ for each of the 
  galaxy clusters in our sample. The grey band indicates the overall sample
  mean $\langle$\vlos$\rangle$ of $430 \pm 210$~\kms. Red bands denote galaxy clusters identified
  as having a merger along the LOS based on previous analyses and blue bands denote galaxy clusters
  identified as POS mergers or relaxed.}
  \label{fig:vpec}
\end{figure}

The typical per-cluster uncertainty on the value of \vlos\ we obtained
from these fits was 500--1000 \kms, which was a factor of 2--4
larger than the typical expected \vlos\ 
(\eg, \citealt{Evrard2002, Hernandez2010, Nagai2013}).
Not surprisingly, given these uncertainties,
we did not detect a significant non-zero value of
\vlos\ for any single galaxy cluster.
To characterize the galaxy cluster ensemble as a whole, we computed
the inverse variance weighted sample mean 
$\langle$\vlos$\rangle = 430 \pm 210$~\kms.
However, this simple calculation did not account for the intrinsic
cosmological variation in the value of \vlos, and so we also computed
the sample average velocity using a more sophisticated fit based
on the \textsc{linmix\_err} formalism of \citet{Kelly2007}.
From these fits, we obtained a sample average velocity of
$\langle$\vlos$\rangle = 460 \pm 300$~\kms\ and an intrinsic scatter with an rms of
$\sigma_{\textrm{\small int}} = 470 \pm 340$~\kms.

\begin{figure*}
  \centering
  \includegraphics[width=0.35\textwidth]{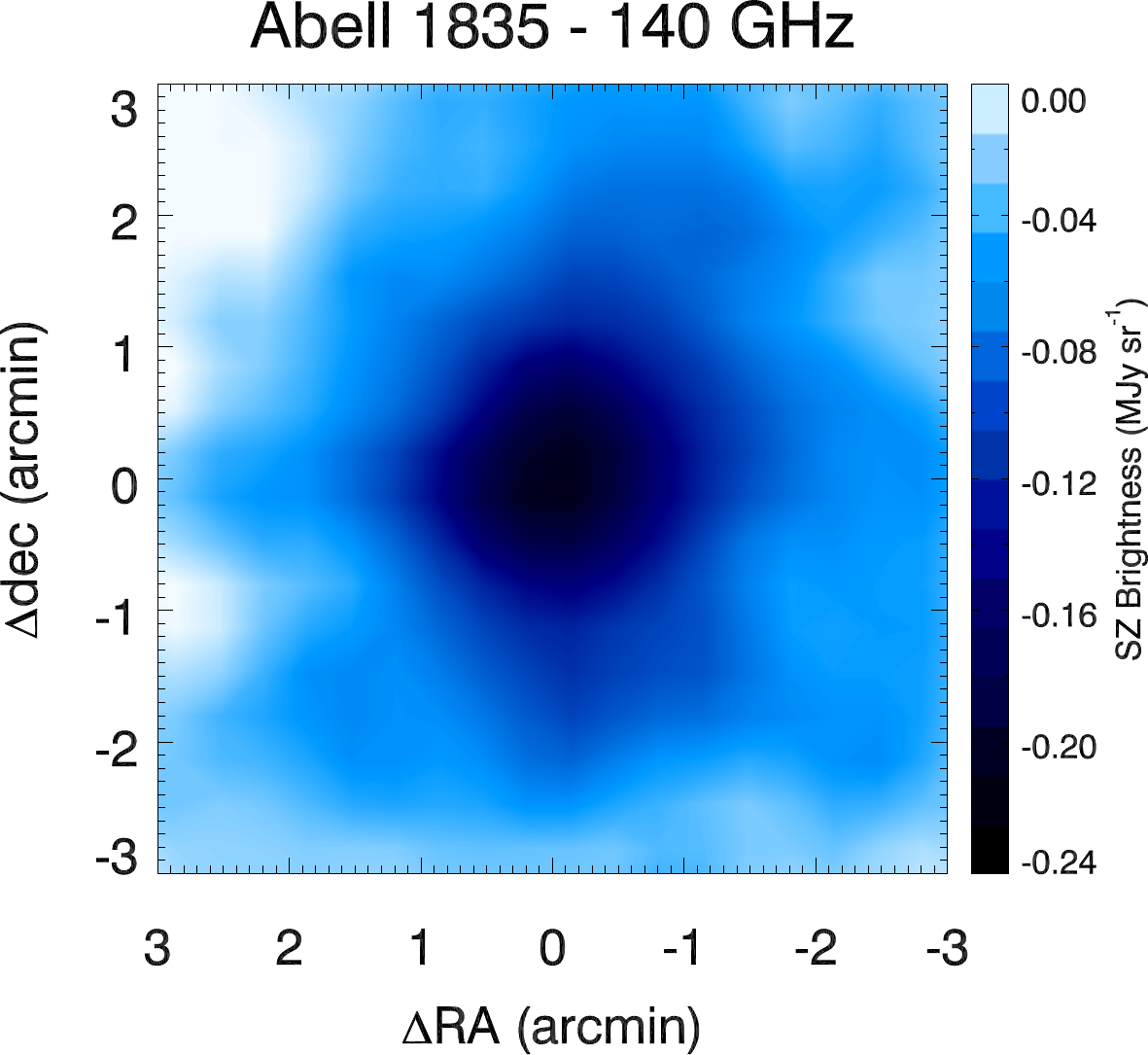}
  \hspace{0.05\textwidth}
  \includegraphics[width=0.35\textwidth]{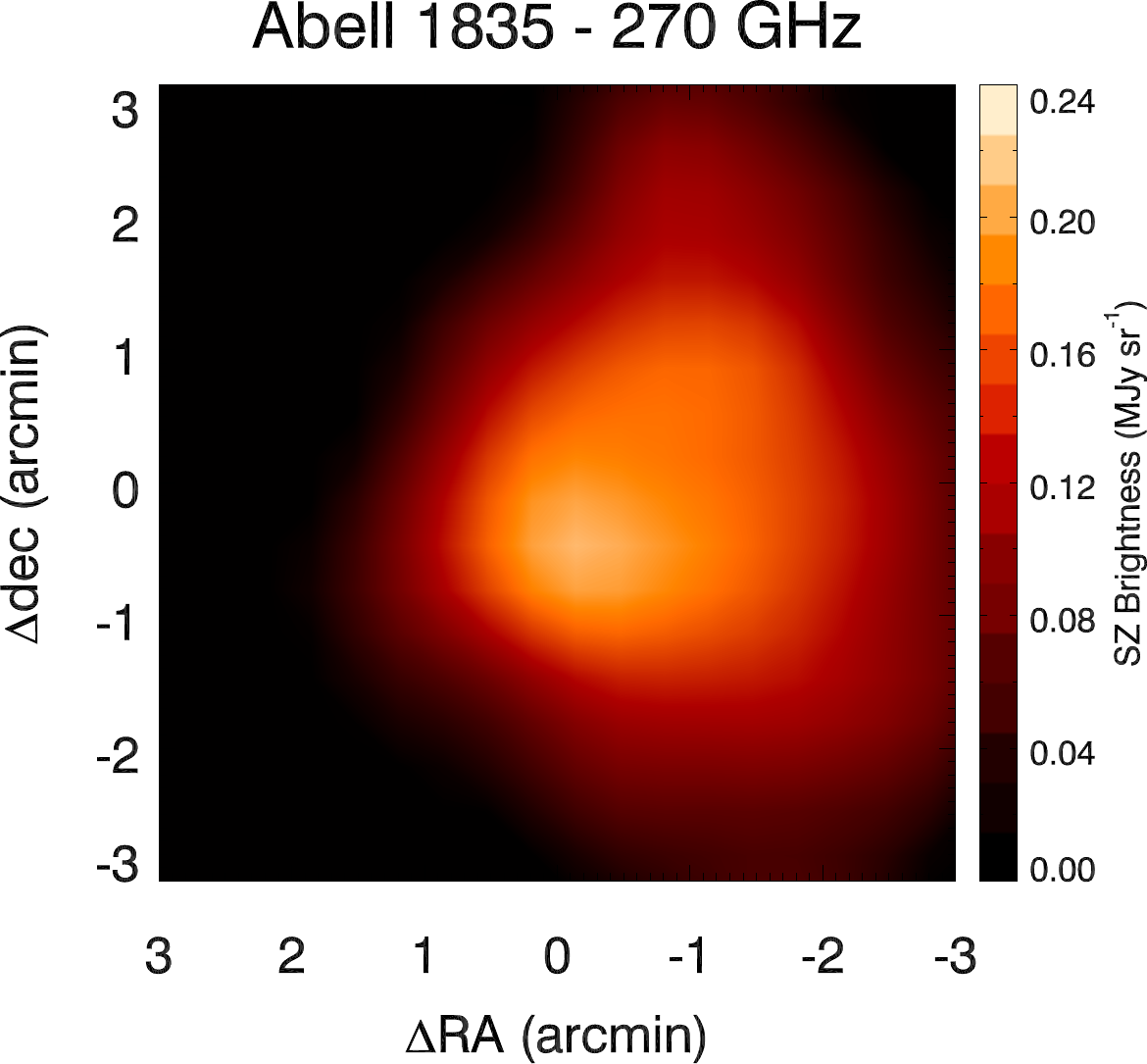}
  \caption{Example SZ effect brightness images at 140 and 270 GHz
  for \aeight. As detailed in the text, bright radio galaxies
  and DSFGs have been subtracted from these images, and they
  have been corrected for the CIB deficit caused by the combination
  of gravitational lensing and bright source subtraction. In addition, the
  filtering effects of our data processing have been deconvolved,
  and the images have been smoothed to a common angular
  resolution of 70\arcsec\ FWHM.
  In the limit of a purely thermal SZ effect signal with
  constant \te, the morphology of the SZ effect brightness
  at both observing frequencies would be the same.
  Differences in shape are indicative of the presence
  of a non-negligible kinematic SZ effect signal and/or
  \te\ variations that produce different relativistic
  corrections to the thermal SZ effect signal over the galaxy cluster face.
  However, in the case of \aeight, the slight differences in morphology
  shown above are fully consistent with noise fluctuations.}
  \label{fig:SZ_images}
\end{figure*}

As expected, the mean velocity we obtained for our sample from both methods
was consistent with zero, although the weighted mean differed at a significance of
$\simeq 2\sigma$. While our uncertainty on the mean velocity was
better than the pioneering measurements from SuZIE \citep{Benson2003}
and the value of $\pm 383$~\kms\
obtained by \citet{Lindner2015} for a similar analysis of eleven galaxy clusters
using data from the ACT and LABOCA, it was notably larger than the value
of $\pm 60$~\kms\ obtained from a \planck\ analysis of $\sim 1750$ 
X-ray-selected galaxy clusters \citep{Planck2014_XIII}.

Our best-fit value for the intrinsic cluster-to-cluster scatter
was consistent with the simulation-based expectation
of $\sim 250$~\kms\ (\eg, \citealt{Evrard2002, Hernandez2010, Nagai2013}),
although with a somewhat large uncertainty of $\pm 340$~\kms.
However, we note that this uncertainty was
comparable to what was obtained from \planck-based 
analyses of large samples of X-ray-selected galaxy clusters (\ie, $<800$~\kms\ at 95\%
confidence in \citealt{Planck2014_XIII} and $350 \pm 270$~\kms\ in
\citealt{Planck2018_LIII}) and slightly better than the upper
limit of 1450~\kms\ obtained by \citet{Lindner2015}.

\subsection{Potential Sources of Bias}

We note that the value of \vlos\ we obtained represents the average
LOS velocity within \radius. However, internal velocities in the ICM
are expected to be comparable to the overall galaxy cluster peculiar velocity,
even when the galaxy cluster is relatively relaxed.
On average, these internal motions were not expected to produce a bias
in the measured value of \vlos, although they were expected to introduce
an rms dispersion of $\simeq 50$--100~\kms, depending on the orientation
and the dynamical state of the galaxy cluster \citep{Nagai2003}.
This dispersion is roughly one order of magnitude below our typical
measurement uncertainty per cluster, and was therefore not included
in our analysis.

The galaxy clusters in our sample are 
not isothermal, and so, in general, our assumption of an isothermal ICM 
produced some slight biases in our results
(see, \eg, \citealt{Chluba2013}).
Because the relativistic corrections to the SZ effect signal
are non-linear with respect to \te, the signal from an isothermal
galaxy cluster will not in general be equal to the signal from
a non-isothermal galaxy cluster with the same mean \te. To estimate
the potential bias from this effect, we computed the the expected SZ effect signal
within \radius\ for the least isothermal galaxy cluster in our sample,
\mseven, using both the isothermal assumption and the 34 different
values of \te\ within the separate {\tt contbin} regions for that cluster.
Even with \te\ ranging from 2 to 24~keV within those separate {\tt contbin} 
regions, the fractional difference
between the SZ effect signals computed using the two methods
was only 0.2\% at 140~GHz and 0.7\% at 270~GHz.
Assuming a similar \te\ structure along the LOS, this calculation indicates
that the potential bias from our isothermal assumption 
was $\lesssim 1$\% for all of the galaxy clusters
in our sample. 

Another, potentially larger source of bias was due to our
use of X-ray spectroscopy from \chandra\ to determine the values of \te.
We note that the thermal SZ effect signal, and relativistic
corrections to the SZ effect signals, depend on the LOS mass-weighted
value of \te. 
Within \radius, hydrodynamical simulations indicate that
the value of \te\ inferred from fitting an X-ray spectrum with a thermal
emission model typically differ from the LOS mass-weighted \te\ at
the level of 4--7\% (\citealt{Nagai2007_2}; see also \citealt{Rasia2014}).
We did not attempt to correct for this difference in our analysis,
although we note that it was sub-dominant compared to our assumed
X-ray calibration uncertainty of 10\%.

In addition, the well established difference in calibration
between the two great X-ray observatories, \chandra\ and \xmm,
may also suggest a potential bias in our results.
For the clusters in our sample, with \te\ generally
between 5--10~keV, \chandra\ has been shown to systematically
measure \te\ values $\simeq 10$--20\% higher 
than \xmm\
(\eg, \citealt{Reese2010, Mahdavi2013, Donahue2014, Schellenberger2015, Madsen2017}).
While it is not clear which observatory has the more accurate
calibration, this difference implies calibration uncertainties
that may exceed the 10\% rms we assumed in our analysis.
Reconciling the \chandra/\xmm\ calibration was beyond the scope of this
work, but a relatively accurate {\it post facto} correction can
be applied to our results if future work is able to better 
determine the effective area of {\it Chandra}. Because the relativistic corrections to 
the SZ effect signals were relatively small for our data
(\eg, $\sim 10$\% changes in \te\ would result in $\sim 1$\%
changes to the relativistic corrections), the spectral shapes
of the thermal and kinematic SZ effect signals will remain nearly
identical for small changes in \te. Therefore,
the thermal and kinematic SZ effect brightnesses obtained from
our analysis would remain largely unchanged. As a result, if the value of
\te\ changes by a factor of $1+\delta_T$, then, to good approximation,
the value of \vlos\
will also change by a factor $1+\delta_T$ and the value \taue\ will change by
a factor of $1/(1+\delta_T)$.

\newpage

\section{Resolved SZ Effect Imaging}
\label{sec:imaging}

\begin{figure*}
  \centering
  \includegraphics[width=\textwidth]{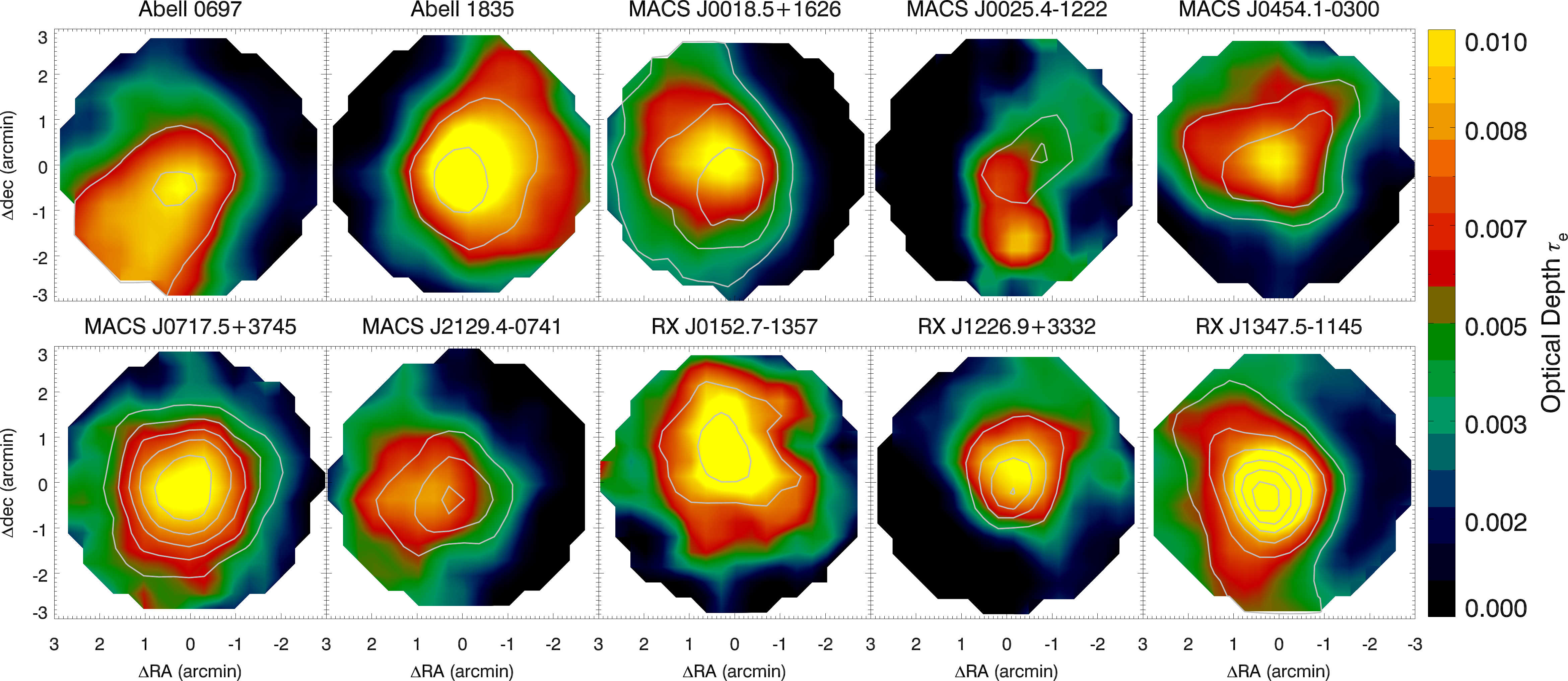}
  \caption{Maps of the electron optical depth \taue\ obtained from our analysis.
    In all cases the images have been smoothed to an effective resolution of 
    70\arcsec\ FWHM, and the grey contours begin at $+3\sigma$ and are separated
    by $2\sigma$. Because the S/N scales mainly with the strength of the thermal
    SZ effect signal, which is the product of \taue\ and \te, the contours
    do not strictly follow the values of \taue\
    due to variations in \te\ over the galaxy clusters' faces.
  }
  \label{fig:tSZ_images}
\end{figure*}

\begin{figure*}
  \centering
  \includegraphics[width=\textwidth]{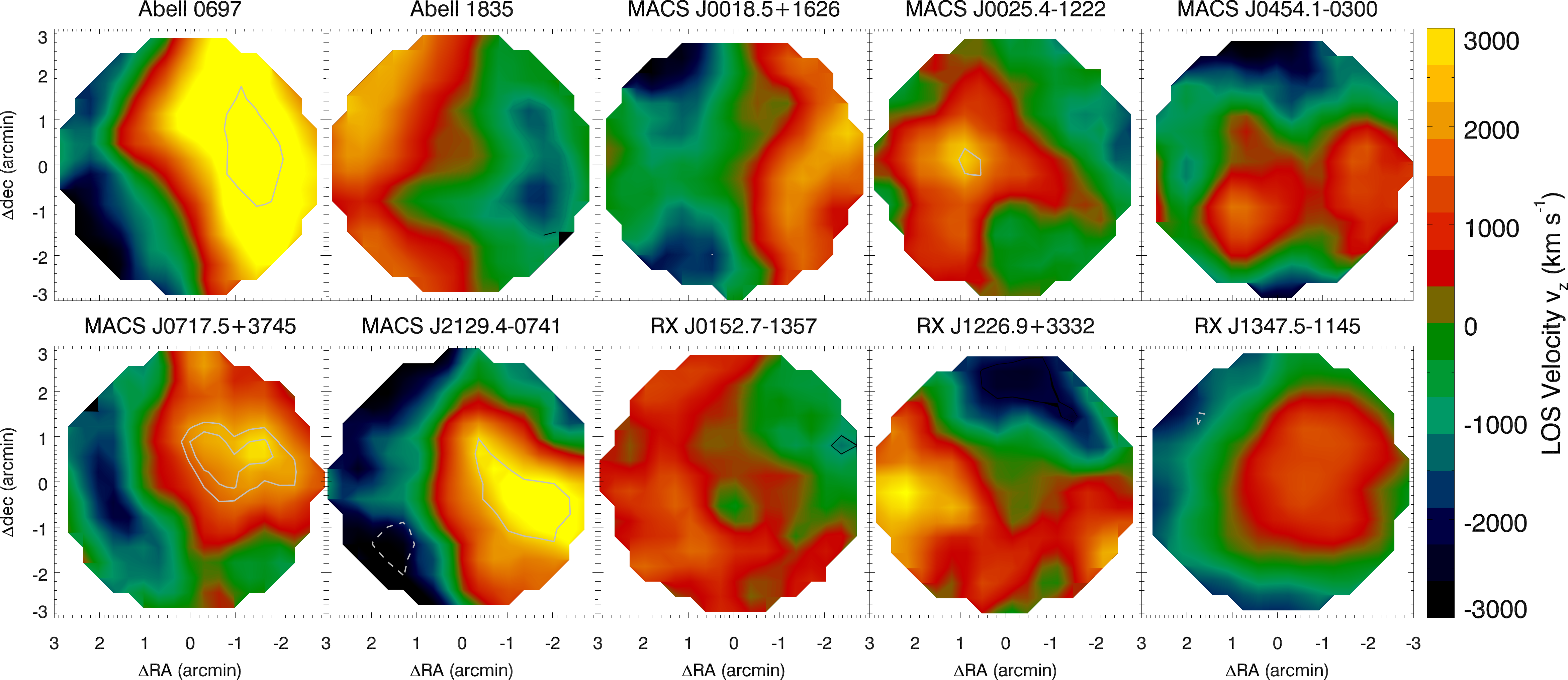}
  \caption{Maps of the LOS velocity \vlos\ obtained from our analysis.
    In all cases the images have been smoothed to an effective resolution of 
    70\arcsec\ FWHM, and the solid/dashed grey contours begin at 
    $+2\sigma$/$-2\sigma$ and are separated
    by $1\sigma$. Because the S/N scales mainly with the strength of the
    kinematic SZ effect signal, which is the product of \vlos\ and \taue,
    the contours do not strictly follow the values of \vlos\
    due to variations in \taue\ over the galaxy clusters' faces.
    Furthermore, to eliminate large un-physical values of \vlos, these
    images have been appodized in regions where the value of \taue\ is less
    than 0.5 times its peak value for each galaxy cluster.
    The only significant detection of \vlos\ for a single sub-structure
    is to the NW of the cluster center of \mseven. This detection is coincident
    with a known merging sub-cluster with a LOS velocity of $\simeq +3000$~\kms.
    While we were not able to detect a non-zero \vlos\ towards a single sub-structure
    in any of the other clusters, we were able to detect an excess variance
    in \vlos\ over the cluster face of \mseven\ at high significance,
    along with lower significance excess \vlos\ variances over the cluster faces of
    \asix, \moone, and \mtwo. See
    Figure~\ref{fig:ksz_rms}.
  }
  \label{fig:kSZ_images}
\end{figure*}

As detailed in \citet{Czakon2015}, it is possible to deconvolve the filtering
effects described in Section~\ref{sec:SZ_red} to obtain an unbiased image 
of the galaxy cluster SZ effect signal. One subtlety is that the filtering completely removes
the mean signal level of the image, and so it must be determined using an
independent measurement. In general agreement with the procedure of 
\citet{Czakon2015}, we used the elliptical gNFW fits from Section~\ref{sec:bulk}
in order to determine the mean signal level of the unfiltered images.
However, one important difference in this work was the addition of 
\planck\ \ymap\ data in constraining the gNFW fits, as it was far
more sensitive to the large angular scale SZ effect signal than
the Bolocam data.
Specifically, for this analysis
we added a constant signal separately to the 140 and 270~GHz 
unfiltered images such that the average surface brightness within \radius\ was
equal to the value obtained from the gNFW model fit. Example images
are shown in Figure~\ref{fig:SZ_images}.

After we obtained these mean-corrected unbiased images, we then convolved them
with a Gaussian kernel to obtain a common resolution of 70\arcsec\ FWHM.
While it would have been possible to use a resolution of 59\arcsec\ FWHM, 70\arcsec\
was chosen as a reasonable compromise between retaining spatial fidelity and 
filtering noise on small angular scales. 
From these images, we then fitted an SZ effect spectrum to each map pixel using
the same procedure applied to the bulk galaxy cluster 
fits that were described in Section~\ref{sec:bulk}. Resolved maps
of \te\ using \chandra\ X-ray spectroscopy were used to estimate 
\te\ within each pixel.
From these fits, we then reconstructed resolved images of the
thermal and kinematic SZ effect signals, which were then combined 
with the \te\ map to obtain images of the electron optical
depth \taue\ (see Figure~\ref{fig:tSZ_images}) and the LOS
velocity \vlos\ (see Figure~\ref{fig:kSZ_images}).

\begin{figure}
  \centering
  \includegraphics[width=0.4\textwidth]{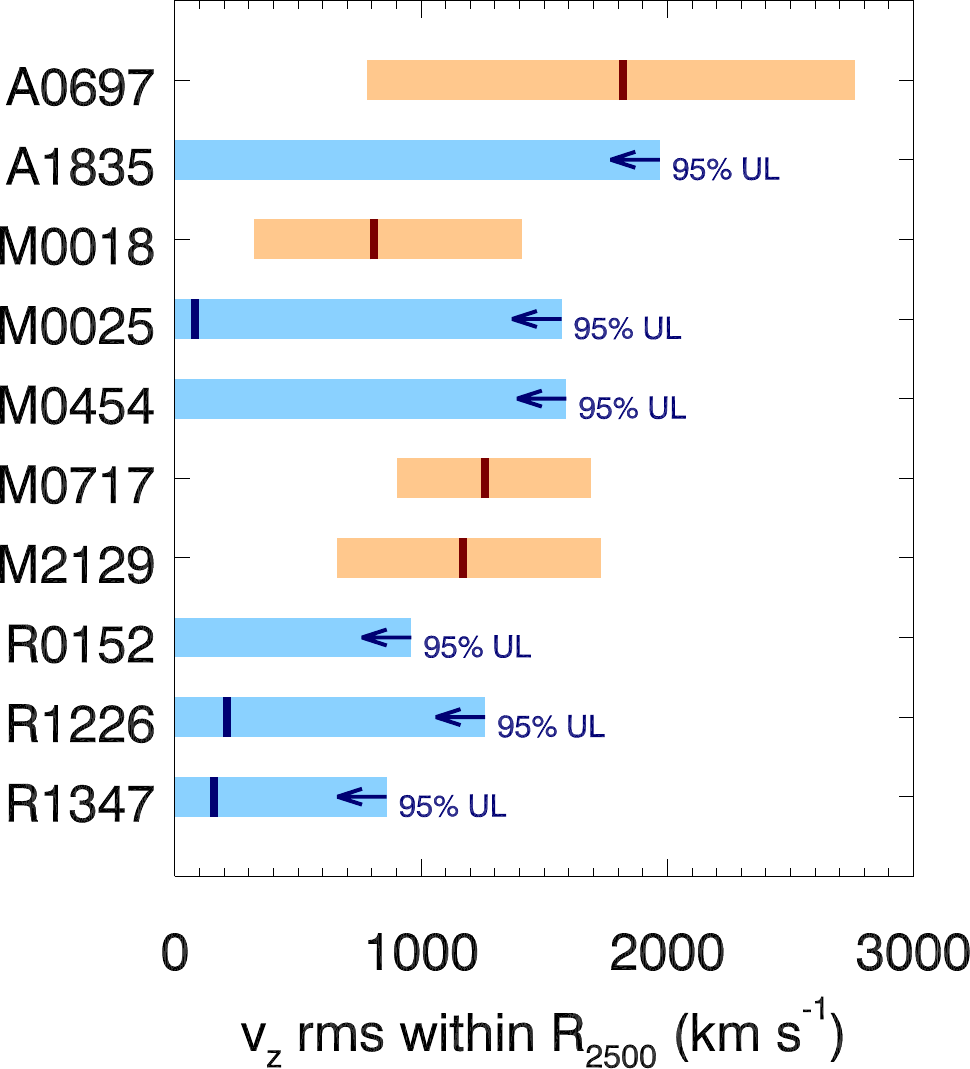}
  \caption{The measured \vlos\ rms within \radius. The solid
    vertical lines represent the best-fit rms value for
    each cluster from the resolved \vlos\ map, with
    $\sigma^2_{vz} = \sigma^2_{map} - \langle \sigma_{noise} \rangle^2$
    to account for the expected rms due to noise fluctuations.
    For three of the clusters, $\sigma_{map} < \langle \sigma_{noise} \rangle$,
    and so no vertical line is shown. The four clusters
    previously identified as LOS mergers are shown in red,
    and the six clusters previously identified as POS mergers
    or relaxed are shown in blue. All four of the LOS mergers
    have an rms $\gtrsim 2\sigma$ from 0, while all six of
    the POS mergers or relaxed clusters have an rms 
    consistent with zero.
    For the clusters with a non-zero detection of the rms,
    the shaded band shows the 68\% confidence region.
    For the clusters with an rms consistent with 0, the
    shaded band extends to the 95\% confidence level upper limit.}
  \label{fig:ksz_rms}
\end{figure}

For all ten galaxy clusters in our sample, the optical depth was imaged at high
significance, with a peak S/N of more than 5. However, the most
significant excursion identified in any of the velocity images
had a S/N of 3, and it was coincident with the merging sub-cluster
in \mseven\ previous described in \citet{Sayers2013}.
Therefore, in nine of the ten clusters we were unable to
detect the LOS velocity of any single sub-structure.
To further search for evidence of underlying LOS velocity sub-structure
below our detection limit within any single resolution element, 
we also computed the rms of the \vlos\ map over the galaxy cluster
face within \radius, $\sigma_{map}$. We then computed an identical rms
from each of the 1000 noise realizations for each cluster ($\sigma_{noise}$), 
which provided an estimate of the expected rms in the absence of
any underlying LOS velocity variations.
We estimated the true internal \vlos\ rms as the difference
between the measured rms and the expected rms due to noise
(\ie, $\sigma^2_{vz} = \sigma^2_{map} - \langle \sigma_{noise} \rangle^2$).
The distribution of $\sigma_{noise}$ values was also used to empirically
determine confidence regions for the value of $\sigma_{vz}$.
The resulting constraints on the rms of \vlos\ within \radius\
for each galaxy cluster are given in Table~\ref{tab:results}
and plotted in Figure~\ref{fig:ksz_rms}. 

All four of the clusters previously identified as likely LOS mergers
had a non-zero measured $\sigma_{vz}$ (at a significance of 
$\simeq 2\sigma$ for \asix, \moone, and \mtwo\ and at a significance
of $\simeq 4\sigma$ for \mseven).
The inferred \vlos\ rms for these clusters was $\gtrsim 1000$~\kms,
$\simeq 3$ times higher than expected from simulations of similar
mass clusters (\eg, \citealt{Nagai2013}).
While of modest statistical significance, our measurements were
therefore consistent with a scenario where each of these four
clusters is undergoing a merger along the LOS, which would boost
the value of $\sigma_{vz}$.
In contrast, the six clusters previous identified as likely
POS mergers or relaxed all had a measured $\sigma_{vz}$
consistent with zero. At a confidence level of 95\%,
the \vlos\ rms for these clusters was $\lesssim 1000$--1500~\kms.
Based on the previously inferred merger geometry for the ten
clusters in our sample, our SZ effect measurements were therefore
able to distinguish LOS mergers from POS mergers and
relaxed clusters.

One of the galaxy clusters in our sample, \mseven, has been the target
of several previous kinematic SZ effect studies, most notably by \citet{Sayers2013}
and \citet{Adam2017}. \citet{Sayers2013} used nearly identical data
to those used in our study, although they included X-ray observations
from {\it XMM-Newton} and they did not use \planck\ SZ effect data.
They also used a much more individualized SZ effect analysis based on
a spatial template derived from the X-ray data and a focus solely on
the signal within 60\arcsec\ diameter apertures centered on sub-clusters
``B'' and ``C''. Their ``direct integration'' results are the most
comparable to those obtained in our more general SZ effect analysis,
and they obtained best-fit \vlos\ values of $+2550 \pm 1050$~\kms\ towards
``B'' and $-500 \pm 1600$~\kms\ towards ``C''. At the same positions
in our \vlos\ map, we obtained values of $2100 \pm 700$~\kms\ and 
$-400 \pm 800$~\kms. The shift to a smaller positive \vlos\ for
sub-cluster ``B'' was driven mainly by our correction for the lensing-induced
deficit in the CIB, which was not included in the analysis of 
\citet{Sayers2013}. This also drove a shift towards a larger negative
\vlos\ for ``C'', but this shift was more than compensated for by the
significantly lower $T_e$ obtained in our analysis, which resulted
in a larger best-fit \taue\ and subsequently smaller magnitude for \vlos.
The smaller uncertainties obtained in our analysis were due
to the combination of: a larger aperture (70\arcsec\ compared to 60\arcsec);
larger best-fit values for \taue, particularly for ``C'';
improved calibration;
and, most significantly, the inclusion of \planck\ SZ effect data to better constrain the
large angular scale signal.

Using completely independent SZ effect measurements, \citet{Adam2017}
measured best-fit \vlos\ values of $+6600^{+3200}_{-2400}$~\kms\ and 
$-4100^{+1600}_{-1100}$~\kms\ for sub-clusters ``B'' and ``C''.
These values are in modest ($\lesssim 2\sigma$) tension with the values
we obtained in our analysis, although we note that \citet{Adam2017}
found \vlos\ equal to $+2100^{+500}_{-450}$~\kms\ for sub-cluster ``B''
using an alternate analysis which included stronger X-ray priors,
fully consistent with our measurement.
Furthermore, the NIKA SZ effect observations used by \citet{Adam2017}
had a factor of $\simeq 3$ finer angular resolution compared to
our Bolocam/AzTEC data, better isolating the sub-clusters and producing
more significant excursions in their \vlos\ map of the galaxy cluster
($5.1\sigma$ and $3.4\sigma$ for ``B'' and ``C'').

\section{Summary}
\label{sec:summary}

We have used observations from Bolocam and AzTEC to image the SZ effect
signal towards a sample of ten galaxy clusters at 140 and 270~GHz. In support
of these data, we have also made use of a number of additional
observations. The \planck\ all-sky \ymap s were used to help constrain
the large-angular scale signal in order to obtain spatial templates of the 
SZ effect signal. In addition, three-band \spire\ imaging was used to subtract
the emission from DSFGs, which was significant compared to the SZ
effect signal at 270~GHz. Furthermore, \hst\ data were used to obtain detailed
mass models for each galaxy cluster in order to properly account for 
lensing of the background CIB. Finally, \chandra\ X-ray spectroscopic imaging
was used to obtain resolved maps of the ICM temperature \te.

From this analysis, we produced galaxy cluster-averaged fits to the SZ effect
brightness at 140 and 270~GHz in order to constrain the average
optical depth and bulk LOS velocity \vlos\ within \radius. Our typical
measurement uncertainties on \vlos\ were 500--1000~\kms, a factor
of 2--4 larger than the typical values of \vlos\ expected from simulations. 
We did not detect \vlos\
at high significance in any single galaxy cluster, and the ensemble average
velocity was consistent with zero, particularly when intrinsic scatter
was accounted for in the fit.
When fitting for the intrinsic scatter, we did not obtain a significant
detection, but we did find an upper limit competitive with those
produced by statistical stacks in CMB survey data.

In addition to fitting for the galaxy cluster-average SZ effect brightness, we also
produced images of the electron optical depth \taue\ and the 
LOS velocity \vlos\ with a resolution of 70\arcsec.
In all cases, \taue\ was detected at high significance near 
the galaxy cluster center.
We did not obtain a significant detection of \vlos\
within any single resolution element for
any of the galaxy clusters in our sample, with the exception
of the previously identified sub-component of \mseven.
However, all four of the clusters previously identified
as likely LOS mergers showed a \vlos\ rms
greater than zero at a significance of $\gtrsim 2\sigma$,
with $\sigma_{vz} \gtrsim 1000$~\kms\ for these objects.
This is a factor of $\simeq 3$ above the \vlos\ rms expected
from simulated clusters of similar masses (\eg, \citealt{Nagai2013}),
strongly indicating a boosted $\sigma_{vz}$ due to a LOS
merger.
In contrast, all six of the clusters previously identified
as likely POS mergers or relaxed had $\sigma_{vz}$
consistent with zero and $\sigma_{vz} \lesssim 1000$--1500~\kms\
at a 95\% confidence level.
Based on the previous characterizations of the merger geometries 
for these galaxy clusters, our SZ effect
data were therefore able to distinguish between LOS 
mergers from POS mergers and relaxed clusters.

In addition to the ICM constraints obtained in our analysis, we also 
quantified the potential bias in measuring the SZ effect signal due
to lensing of the background DSFGs that comprise the CIB. When individual
bright DSFGs are identified and subtracted, lensing produces an on-average
deficit in the surface brightness of the CIB. For the galaxy clusters in our sample,
the total surface brightness of this deficit was typically $\simeq 15$\%
of the total surface brightness of the SZ effect signal at 270~GHz, although 
it was as large as 25\% for one galaxy cluster (\mtwo).

In contrast to some other recent kinematic SZ effect analyses (\eg,
\citealt{Sayers2013} and \citealt{Adam2017}), we did not make use
of X-ray data to model the shape of the ICM pressure or density,
although we did make use of resolved temperature maps from
spectroscopic \chandra\ X-ray observations.
While such X-ray density and pressure information 
can be useful in breaking degeneracies in kinematic SZ effect
measurements (\eg, \citealt{Flender2017}), they can be
difficult to include for detailed studies of individual galaxy clusters with
complicated merger geometries, as was the case for nine of the
ten objects in our sample. For example, deprojections, which
assume a spherical geometry, can only be applied in special
cases for merging clusters 
(\eg, it was only used for one ICM sub-component in the analysis of \citealt{Adam2017}).
\citet{Sayers2013} avoided this complication by using the SZ effect
data to constrain the LOS extent of an X-ray derived pseudo-pressure
map of \mseven, although the resulting constraints were only marginally
better than those obtained in our current analysis without
an X-ray template.
We therefore decided for this work not to pursue an analysis based on X-ray
maps of the ICM density or pressure.

Looking to the future, instruments like TolTEC \citep{Bryan2018} will 
be able to provide much deeper SZ effect observations, at finer angular
resolution, and in more observing bands. Scheduled to be installed
in 2019 on the 50 meter Large Millimeter Telescope Alfonso Serrano (LMT)
in M{\'e}xico, TolTEC will simultaneously observe at 150, 220, and 280~GHz,
providing images of the thermal and kinematic SZ effect signals while robustly
detecting (and subtracting) the contaminating signals from DSFGs. Compared to the
data used for this work, TolTEC promises an order of magnitude or more
improvement in achievable map depth, opening up the prospect of high
significance imaging of ICM velocity structures at $\lesssim 10$\arcsec\
resolution. Longer-term concepts, such as AtLAST \citep{Bertoldi2018}
and CSST \citep{Golwala2018}, promise to deliver much larger fields of
view than TolTEC and expanded spectral coverage (\eg, 90--400~GHz), further enhancing
the potential scientific reach of detailed SZ effect studies.

\section{Acknowledgments}

Sunil Golwala, Erik Reese, Jack Sayers, and Grant Wilson were supported by 
NASA/NNX15AV66G. 
This work was partially supported by Consejo Nacional de Ciencia y Technolog{\'i}a
(CONACYT) project FDC-2016-1848.
The scientific results reported in this article are based in part on data
obtained from the \chandra\ Data Archive.
This research has made use of software provided by the \chandra\ X-ray
Center (CXC) in the application packages CIAO, ChiPS, and Sherpa.
We thank Georgiana Ogrean for providing her X-ray reprocessing
and analysis scripts.
We thank the referee for providing several useful suggestions that
have improved this manuscript.

\bibliography{ms}
\bibliographystyle{aasjournal}

\end{document}